%% file: main.tex
\documentclass[%
 reprint,
superscriptaddress,
 amsmath,amssymb,
 aps,
 prx,
floatfix,
]{revtex4-2}

\usepackage[T1]{fontenc}
\usepackage[utf8]{inputenc}
\usepackage{amsmath,graphicx}
\usepackage{amssymb}
\usepackage{epsfig,graphicx}
\usepackage{mathrsfs}
\usepackage[active]{srcltx}
\usepackage{mdframed}
\usepackage{mathtools}
\usepackage{xcolor}
\usepackage{hyperref}
\usepackage{braket}
\usepackage{gensymb}
\usepackage{enumerate}
\usepackage{epsfig,libertine}
\usepackage[libertine]{newtxmath}
\usepackage{layouts}
\usepackage{dcolumn}
\usepackage{bm}
\usepackage{physics}
\usepackage{float}
\makeatletter
\let\newfloat\newfloat@ltx
\makeatother
\usepackage{algorithm}
\usepackage[noend]{algpseudocode}
\usepackage{caption}
\begin{document}

% \preprint{APS/123-QED}

\title{Efficient Quantum Gradient and Higher-order Derivative Estimation \protect \\ via Generalized Hadamard Test}

\author{Dantong Li}
\affiliation{Department of Computer Science, Yale University, New Haven, Connecticut 06511, USA}
\affiliation{Yale Quantum Institute, Yale University, New Haven, Connecticut 06520, USA}
\author{Dikshant Dulal}
\affiliation{Yale-NUS College, 16 College Avenue West, 138527, Singapore}
\affiliation{SoftServe, Inc., 19 Cecil Street, 049704, Singapore}
\author{Mykhailo Ohorodnikov}
\affiliation{SoftServe, Inc., 19 Cecil Street, 049704, Singapore}
\author{Hanrui Wang}
\affiliation{Department of Computer Science, Massachusetts Institute of Technology, Cambridge, MA 02139, USA}
\author{Yongshan Ding}
\affiliation{Department of Computer Science, Yale University, New Haven, Connecticut 06511, USA}
\affiliation{Yale Quantum Institute, Yale University, New Haven, Connecticut 06520, USA}
\affiliation{Department of Applied Physics, Yale University, New Haven, Connecticut 06520, USA}
%%%%%%%%%%%%%%%%%%%%%%%%%%%%%%%%%%%%
%%%%%%%%-----NEWCOMMAND-----%%%%%%%%%

% Define a command for nested commutators
\newcommand{\blue}[1]{\textcolor{blue}{#1}}
\newcommand{\StatexIndent}[1][3]{%
  \setlength\@tempdima{\algorithmicindent}%
  \Statex\hskip\dimexpr#1\@tempdima\relax}
\makeatother
\newcommand{\remark}[1]{\textbf{Remark:} #1}
\newtheorem{definition}{Definition}

% Define a new command for the += sign
\newcommand{\pluseq}{\mathrel{+}=}

%%%%%%%%%%%%%%%%%%%%%%%%%%%%%%%%%%%%

\date{\today}

\input{0_abstract}
\maketitle
%%%%%% -- PAPER CONTENT STARTS-- %%%%%%%%

\input{1_introduction}
\input{2_background}
\input{3_gradient_circuits}
\input{4_evaluation}
\input{5_discussion}
\input{6_conclusion}

\input{8_acknowledgements}
\appendix
\input{7_appendix}

\bibliography{refs}

\end{document}

%% file: 0_abstract.tex
\begin{abstract}

In the context of Noisy Intermediate-Scale Quantum (NISQ) computing, parameterized quantum circuits (PQCs) represent a promising paradigm for tackling challenges in quantum sensing, optimal control, optimization, and machine learning on near-term quantum hardware. Gradient-based methods are crucial for understanding the behavior of PQCs and have demonstrated substantial advantages in the convergence rates of Variational Quantum Algorithms (VQAs) compared to gradient-free methods. However, existing gradient estimation methods, such as Finite Difference, Parameter Shift Rule, Hadamard Test, and Direct Hadamard Test, often yield suboptimal gradient circuits for certain PQCs.
To address these limitations, we introduce the Flexible Hadamard Test, which, when applied to first-order gradient estimation methods, can invert the roles of ansatz generators and observables. This inversion facilitates the use of measurement optimization techniques to efficiently compute PQC gradients. Additionally, to overcome the exponential cost of evaluating higher-order partial derivatives, we propose the \(k\)-fold Hadamard Test, which computes the \(k^{th}\)-order partial derivative using a single circuit.
Furthermore, we introduce Quantum Automatic Differentiation (QAD), a unified gradient method that adaptively selects the best gradient estimation technique for individual parameters within a PQC. This represents the first implementation, to our knowledge, that departs from the conventional practice of uniformly applying a single method to all parameters.
Through rigorous numerical experiments, we demonstrate the effectiveness of our proposed first-order gradient methods, showing up to an \(O(N)\) factor improvement in circuit execution count for real PQC applications. Our research contributes to the acceleration of VQA computations, offering practical utility in the NISQ era of quantum computing.

\end{abstract}

%% file: 1_introduction.tex
\section{Introduction}
Quantum computers have the potential to revolutionize technology by providing improved solutions to optimization problems in fields such as molecular simulation \cite{grimsley2019adaptive}, drug discovery \cite{jones2019variational}, finance \cite{orus2019quantum}, and machine learning \cite{lloyd2013quantum, rebentrost2014quantum}. A prominent research area that has emerged for utilizing Noisy Intermediate-Scale Quantum (NISQ) devices is Parameterized Quantum Circuits (PQCs). PQCs are crucial in various domains including quantum sensing, quantum optimal control, and quantum metrology. Most importantly, they enable a family of NISQ algorithms called Variational Quantum Algorithms (VQAs), which leverage hybrid systems of quantum and classical computers to find approximate solutions. This hybrid computing paradigm shifts part of the computational burden from quantum to classical processors, making VQAs particularly valuable for near-term devices.

\vspace{0.5em} 
\textbf{Why measuring quantum gradients?}
To understand how parameter changes influence the measurement outcomes of PQCs, it is natural to study their gradients. Specifically, gradients aid in discovering highly entangled states that enhance quantum metrology in noisy environments \cite{koczor2020variational, meyer2021variational} and optimizing parameters in quantum control theory \cite{magann2021pulses, yang2022variational, werschnik2007quantum, li2017hybrid}. Additionally, gradients have been shown, in practice, to accelerate convergence in numerous VQAs, including the Quantum Approximate Optimization Algorithm (QAOA) \cite{QAOA, qaoa_proteinFolding, moll2018quantum}, Variational Quantum Eigensolver (VQE) \cite{VQEs, kandala2017hardware, kokail2019self, mcclean2016theory}, Quantum-Assisted Quantum Compilation (QAQC) \cite{khatri2019quantum, sharma2020noise}, and Quantum Neural Networks (QNNs) \cite{abbas2021power, mcclean2018barren, wittek2014quantum, biamonte2017quantum, schuld2018supervised, benedetti2019parameterized}. Also theoretically, it has been proven that gradient-based optimizations significantly enhance convergence speed compared to gradient-free methods \cite{harrow2021low}, highlighting the demand for precise and efficient gradient estimation techniques.

\begin{figure*}[t]
    \captionsetup{justification=raggedright}
    \centering
    \includegraphics[width=\textwidth]{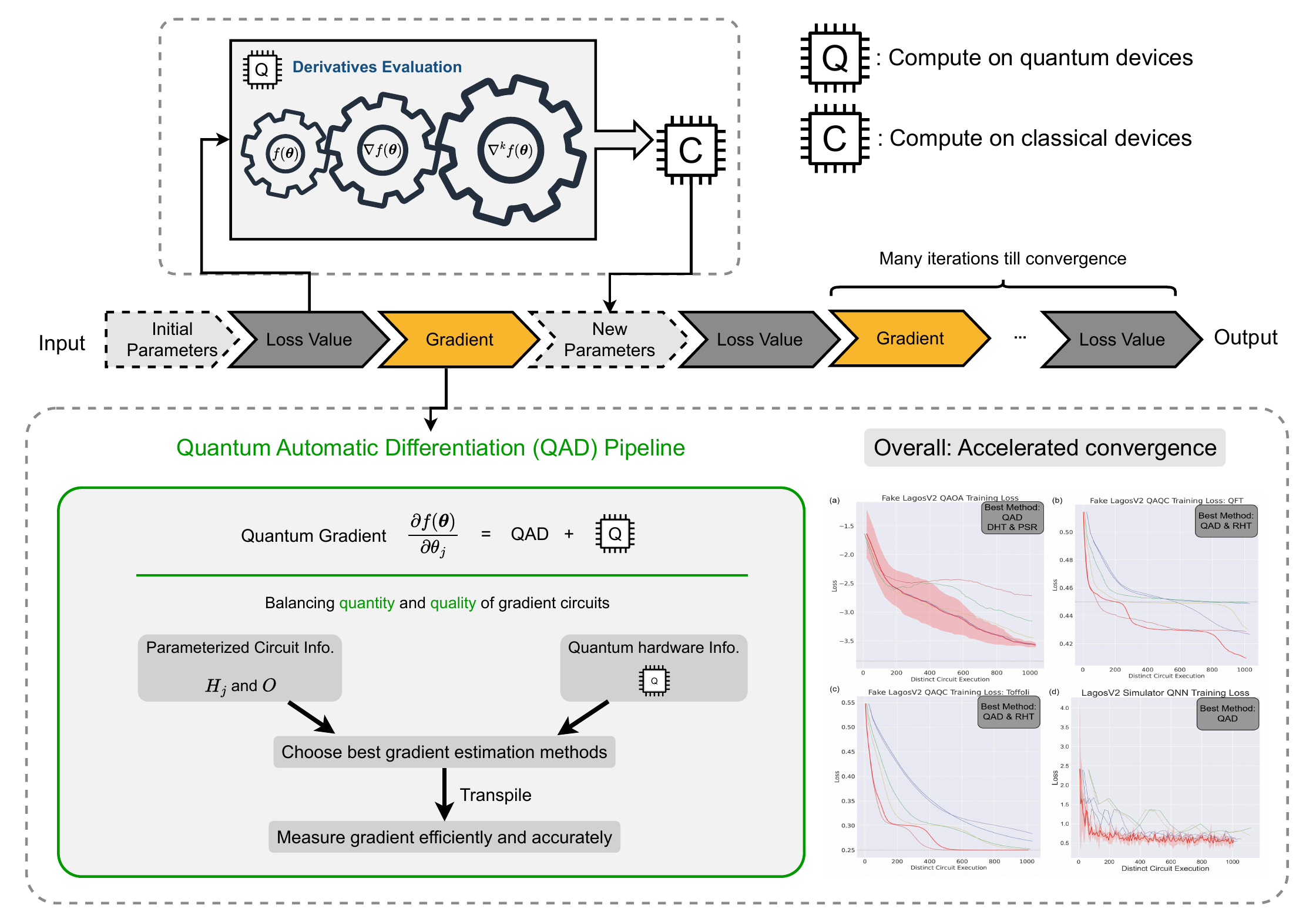}
    \caption{ Parameterized Quantum Circuits Workflow: The optimization process begins with an input state and initial parameters, followed by the execution of a quantum circuit to evaluate the loss function. Gradients and higher-order derivatives are then estimated using quantum computers. These estimates are utilized by a classical computer to update the parameters. This iterative process continues until convergence. During each iteration, the QAD framework selects the optimal gradient methods, balancing efficiency and reliability in gradient estimation. The bottom-right of the diagram features a plot of training performance for QAOA, QAQC, and QNN using various gradient methods, demonstrating that our proposed QAD method achieves superior performance.}
    \label{fig:hybrid}
\end{figure*}

\vspace{0.5em} 
\textbf{Existing Gradient Methods}
Various approaches have been developed for gradient computation. Finite Difference (FD) and Simultaneous Perturbation Stochastic Approximation (SPSA) are commonly employed stochastic approximation methods, where the gradient is estimated by computing differences between function values evaluated at neighboring points in the parameter space. While straightforward, FD and SPSA can suffer from numerical instability and inaccuracy, particularly in high-dimensional and complex landscapes.
The Parameter Shift Rule (PSR) offers an efficient means of calculating the analytical gradient of parameterized quantum gates on quantum hardware. Additionally, the Hadamard Test (HT) gradient method leverages indirect measurements through a Hadamard test, enabling the estimation of quantum gradients for a broader range of parameterized gates compared to PSR.
Later Direct Hadamard Test (DHT) gradient method was proposed to replace indirect measurements with direct ones, facilitating gradient estimation in NISQ systems with limited size and connectivity. DHT presents a qubit-efficient alternative suitable for NISQ era.

\vspace{1em} 
\textbf{Key Challenges}

\begin{itemize}
    \item Classical Backpropagation Intractability: The application of classical backpropagation becomes infeasible when computing gradients of large PQCs. The evaluation of quantum gradients necessitates the utilization of quantum computers, introducing unique computational challenges.
    \item Suboptimal Quantum Gradient Circuits: Existing quantum gradient methods often produce inefficient gradient circuits for certain PQCs, resulting in issues such as deep circuit depth and the extensive need for repeated measurements. Furthermore, all current methods incur an exponential cost when calculating higher-order partial derivatives, highlighting the need for more efficient algorithms tailored to these specific cases.
    \item Need for Specialized Gradient Methods: Different parameters in PQCs may benefit from distinct gradient computation methods. Consequently, a one-size-fits-all solution is not optimal, highlighting the challenge and the importance of developing tailored approaches for individual parameters.
\end{itemize}

\textbf{Key Ideas}
In this study, we introduce two innovative and efficient gradient calculation methods: the Flexible Hadamard Test gradient method and the $k$-fold Hadamard Test gradient method, to address the challenge of suboptimal quantum gradient circuits. The Flexible Hadamard Test gradient method, combined with measurement optimization techniques, reduces circuit execution costs for PQC gradient estimation, potentially expediting the training of VQAs, especially in scenarios where existing methods underperform. Specifically, this general technique can transform existing methods such as the Hadamard Test and Direct Hadamard Test into the Reversed Hadamard Test (RHT) and Reversed Direct Hadamard Test (RDHT) methods, which evaluate gradients by exchanging the roles of unitary gates and measurements. These reversed gradient methods provide additional options for selecting gradient methods for PQCs.
The $k$-fold Hadamard Test gradient method uses a single circuit to estimate the \(k^{th}\)-order partial derivatives in PQCs, compared to existing methods that require \(2^k\) distinct circuits.
To tackle the challenge of different parameters in one PQC potentially benefiting from different gradient methods, we propose the Quantum Automatic Differentiation (QAD) gradient method for the first-order gradient of PQCs. To the best of our knowledge, this is the first implementation of a quantum gradient method that allows tailoring different gradient methods to different parameters in PQCs. We conduct extensive architectural-level analysis and numerical experiments to demonstrate the strengths of the reversed methods and the effectiveness of QAD. Specifically, we provide numerical evidence showcasing the advantages of reversed methods in QAQC, which can save a factor of \(O(N)\) circuit executions for some common use cases. Additionally, we demonstrate the effectiveness of QAD in practical applications through a QNN classification task, where QAD shows a \(9\times\) advantage over the naive parameter shift rule.

\textbf{Core Contributions}
To summarize, our work makes the following contributions:
\begin{itemize}
    \item Introduction of Flexible Hadamard Test: We present a general technique that leads to two novel reversed gradient estimation algorithms, RHT and RDHT.
    \item Introduction of $k$-fold Hadamard Test: We propose an efficient method for estimating higher-order partial derivatives in PQCs using a single circuit, significantly reducing the required number of circuits.
    \item Introduction of Quantum Automatic Differentiation Software: We develop and implement an adaptive gradient method that assigns the best gradient computation technique for individual parameters. To our knowledge, this is the first implementation that allows for assigning different algorithms to different parameters within the same PQC. We also provide a comprehensive architectural-level analysis and demonstrate significant improvements for practical VQA applications.
\end{itemize}
Overall, our work advances the field by providing innovative algorithms for gradient and higher-order derivative computation, as well as an adaptive approach that enhances the performance and efficiency of VQAs. These contributions offer valuable insights and practical solutions across various application domains, showcasing significant improvements in gradient estimation and execution efficiency.

The structure of this paper is as follows: In Section \ref{sec:background}, we describe the foundational concepts of parameterized quantum circuits, their gradient expressions, existing gradient methods, and the technique of measurement grouping. In Section \ref{sec:efficient_grad}, we introduce our proposed efficient gradient estimation algorithms and QAD as an aggregated method. In Section \ref{sec:evaluation}, we conduct numerical experiments to demonstrate the effectiveness of the reversed methods and QAD. Section \ref{sec:related_work} discusses how our work relates to the existing literature and outlines potential future research directions. Finally, in Section \ref{sec:conclusion}, we summarize the contributions of this paper.

\begin{table}[b]
\small
\begin{tabular}{>{\bfseries}l l }
DHT & Direct Hadamard Test \\
FD & Finite Difference \\
HT & Hadamard Test \\
NISQ & Noisy Intermediate-Scale Quantum \\
PQC & Parameterized Quantum Circuit \\
PSR & Parameter Shift Rule \\
QAQC & Quantum-Assisted Quantum Compilation \\
QAD & Quantum Automatic Differentiation \\
QAOA & Quantum Approximate Optimization Algorithm \\
QNN & Quantum Neural Network \\
RDHT & Reversed Direct Hadamard Test \\
RHT & Reversed Hadamard Test \\
SPSA & Simultaneous Perturbation Stochastic Approximation \\
VQA & Variational Quantum Algorithm \\
VQE & Variational Quantum Eigensolver \\
\end{tabular}
\caption{Acronyms used in this paper.}
\label{tab:acronyms}
\end{table}

%% file: 2_background.tex
\section{Background}
\label{sec:background}

\subsection{Parameterized Quantum Circuits}

Parameterized Quantum Circuits (PQCs) are versatile tools in the NISQ era, playing a crucial role in various fields. They are integral to variational quantum algorithms (VQAs) \cite{jones2019variational, VQAs, du2022efficient, VQEs, kandala2017hardware, kokail2019self, mcclean2016theory, QAOA, moll2018quantum, abbas2021power, wittek2014quantum, biamonte2017quantum, schuld2018supervised, benedetti2019parameterized}, quantum sensing \cite{kaubruegger2019variational, koczor2020variational, meyer2021variational}, and optimal control \cite{magann2021pulses, yang2022variational}. Additionally, PQCs are pivotal in the quest for quantum supremacy and advantage \cite{lund2017quantum, harrow2017quantum}. By providing a concrete way to implement algorithms, PQCs leverage quantum computational power to solve problems across various domains, highlighting the potential of quantum computing in both theoretical and practical applications.

Among these, VQA based on PQCs have emerged as a promising candidate to achieve quantum advantage in solving a diverse range of optimization problems. Notable examples include the Variational Quantum Eigensolver (VQE) for molecular simulation in chemistry \cite{VQEs, kandala2017hardware, kokail2019self, mcclean2016theory}, the Quantum Approximate Optimization Algorithm (QAOA) for combinatorial optimization \cite{QAOA, qaoa_proteinFolding, moll2018quantum}, and Quantum Neural Networks (QNNs) for machine learning \cite{abbas2021power, mcclean2018barren, wittek2014quantum, biamonte2017quantum, schuld2018supervised, benedetti2019parameterized}. These problems can be effectively framed as the search for the minimum of a cost function. PQCs begin by selecting a sequence of the parameterized gates as shown below in Fig~\ref{fig:qc_ansatz}.
\begin{equation}
     U_{1:n}(\boldsymbol{\theta}) =  U_n(\theta_n) \cdots U_1(\theta_1) =  e^{-iH_n\theta_n/2} \ldots e^{-iH_1\theta_1/2}\label{eqn:Variational Ansatz}
\end{equation}
where
\begin{align}
    U_j(\theta_j) = e^{-iH_j\theta_j/2}\label{eq:Uj}
\end{align}
$H_j$ is the \textbf{generator} of the gate $U_j$. To simplify notations, we will use \( U_{1:n} \) and \( U_j \) instead of \( U_{1:n}(\boldsymbol{\theta}) \) and \( U_j(\theta_j) \) from this point forward. Two examples of parameterized gates include single-qubit Rotation-Z gate and two-qubit Rotation-ZZ gate:
\begin{align*}
    R_Z(\theta) &= e^{-i\sigma_Z\theta/2} \\
    R_{ZZ}(\theta) &= e^{-i\sigma_Z\otimes\sigma_Z\theta/2}
\end{align*}

\begin{figure}[h]
    \captionsetup{justification=raggedright}
    \centering
    \includegraphics[width=\linewidth]{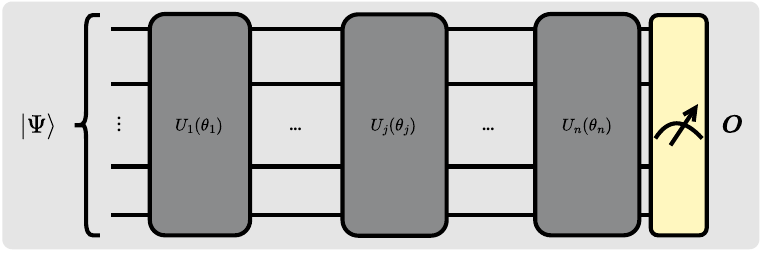}
    \caption{Parameterized Quantum Circuit: input state denoted as $\ket{\Psi}$; quantum gates are parametrized by $\{\theta_j\}_j$; and the quantum system is measured against observable $O$}
    \label{fig:qc_ansatz}
\end{figure}

The gates are then applied to a prepared quantum state $\ket{\Psi}$, which is commonly chosen to be $\ket{0}^{\otimes N}$ and the resulting quantum state is measured with some observable $O$, as illustrated in Fig~\ref{fig:qc_ansatz}. The expected value of the measurement outcome is the cost function $f(\boldsymbol{\theta})$, which parameterizes the solution to the problem in hand.  It is an implicit assumption of PQCs that the cost function cannot be efficiently computed by classical computers but is easy to evaluate with a quantum circuit\cite{VQAs}.  Concretely $f(\boldsymbol{\theta})$ is given by 
\begin{align}
    f(\boldsymbol{\theta}) = \bra{\Psi}U_{1:n}^\dagger O U_{1:n}\ket{\Psi} = \bra{\boldsymbol{\theta}}O\ket{\boldsymbol{\theta}}\label{eq:f(theta)}
\end{align}
where the parameterized quantum state after evolution is $\ket{\boldsymbol{\theta}} = \ket{\theta_1, \theta_2, \ldots \theta_n}= U_{1:n}\ket{\Psi}$. 

\begin{remark}
    Note that while the input quantum states can generally be mixed states rather than pure states, for simplicity and to align with conventional notations, we restrict our discussion to pure states in this paper. However, our results can be easily generalized to mixed states.
\end{remark}

\subsection{PQC Gradients}
Let us first derive the analytical expression for the gradient of $f(\boldsymbol{\theta})$ in Eq.~\eqref{eq:f(theta)} with respect to the $j^{\text{th}}$ parameter $\theta_j$, 

\begin{align}
    \pdv{f(\boldsymbol{\theta})}{\theta_j} 
    &= -\Im{\bra{\Psi} U_{1:j}^\dagger H_j U_{(j+1):n}^\dagger O U_{1:n} \ket{\Psi}} \label{eq:grad_def_HT} \\
    &= \frac{i}{2} \bra{\boldsymbol{\theta}} [U_{(j+1):n} H_j U_{(j+1):n}^\dagger, O] \ket{\boldsymbol{\theta}} \label{eq:grad_def_com}
\end{align}

Gradient-based optimization techniques require accurate estimation of the values in Eq.~\eqref{eq:grad_def_HT} and Eq.~\eqref{eq:grad_def_com}. Although these equations compute the same value, they serve as the basis for different gradient estimation algorithms, as we will explain in the next section. Quantum computers are needed to evaluate such expressions when the size of quantum systems becomes large and intractable for classical computers. In the following, we will introduce the existing gradient estimation methods using quantum hardwares.

\subsection{Quantum Gradient Methods}
\label{sec:bg_gradient_methods}

Evaluating Eq.~\eqref{eq:grad_def_HT} or Eq.~\eqref{eq:grad_def_com} on quantum computers necessitates ingenious approaches, as the operators $H_j$ and $O$ are incorporated asymmetrically within the formula. Current methods are either tailored for specific scenarios or incur additional costs to accommodate more general applications.

\subsubsection{Parameter Shift Rule (PSR)}
\label{sec:PSR}
PSR\cite{vanilla_PSR, crooks_PSR, fourtermPSR, stochasticPSR, generalPSR, mitarai2018quantum, mari2021estimating} is an efficient but restricted method for computing gradients for PQCs. PSR works only when the generator $H_j$ in Eq.~\eqref{eq:Uj} has only two distinct eigenvalues. In that case, the gradient expression in Eq.~\eqref{eq:grad_def_HT} and Eq.~\eqref{eq:grad_def_com} can be written as follows:

\begin{equation}
\frac{\partial f(\boldsymbol{\theta})}{\partial \theta_j} =
c \left[f_{\boldsymbol{\theta}_{-j}}\left(\theta_j + \frac{\pi}{4c}\right) - f_{\boldsymbol{\theta}_{-j}}\left(\theta_j - \frac{\pi}{4c} \right) \right],\label{eq:ParameterShiftRule}
\end{equation}
where the coefficient \(c\) is defined as \(\frac{1}{4}(h_2 - h_1)\). Here, \(h_1\) and \(h_2\) represent the distinct eigenvalues of the generator \(H_j\) associated with the unitary operation \(U_j\). The notation \(f_{\boldsymbol{\theta}_{-j}}\) refers to the cost function evaluated with all parameters except the \(j\)-th one held constant:
\begin{align*}
    f_{\boldsymbol{\theta}_{-j}}(x) = f(\theta_1, \cdots, \theta_{j-1}, x, \theta_{j+1}, \cdots, \theta_n)
\end{align*}

As illustrated in Fig~\ref{fig:grad_methods}(a), PSR only needs two execution of the cost function $f$ for the gradient. PSR is similar to Finite Difference but does not require infinitely small changes of the parameter value for unbiased gradient. PSR can be extended for more complex gates through their decomposition, as outlined in \cite{crooks_PSR}. However, the computational cost grows exponentially with the complexity of the gate decomposition.

\subsubsection{Hadamard Test (HT)}
\label{sec:HT}
HT \cite{guerreschi2017practical, ekert2002direct, li2017efficient, romero2018strategies} is naturally used to estimate Eq.~\eqref{eq:grad_def_HT} treating the whole middle part between $\bra{\Psi} U_{1:j}^\dagger$ and $U_{1:j} \ket{\Psi}$ as a huge observable. HT can calculate the gradient for any generator $H_j$ in parameterized gate $U_j$ in Eq.~\eqref{eq:Uj}. Since any Hermitian matrix can be decomposed into a linear combination of Pauli words \cite{nielsen2002quantum}, we denote the pauli expansion of $H_j$ as:

\begin{equation}
    H_j = \sum\limits_{k=1}^{N_j}\beta_k^{(j)} Q_k^{(j)}.
    \label{eq:hj_decomposition}
\end{equation}

For each \(Q_k^{(j)}\) in the decomposition, the method utilizes a Hadamard test circuit, requiring an additional ancillary qubit, as illustrated in Fig.~\ref{fig:grad_methods}(b), to evaluate:
\begin{align}
    \langle X_0\otimes O \rangle_k =  -\Im{\bra{\Psi} U_{1:j}^\dagger Q_k^{(j)} U_{(j+1):n}^\dagger O U_{1:n} \ket{\Psi}}
\end{align}
where $X_0\otimes O$ is the observable for the Hadamard test circuit. This results in a total of \(N_j\) such circuits executed to obtain the gradient values. The $\{\beta_k^{(j)}\}_k$ weighted sum of the $N_j$ expected values of the measurement from the gradient circuits is exactly $\frac{\partial f(\boldsymbol{\theta})}{\partial \theta_j}$:

\begin{equation}
    \frac{\partial f(\boldsymbol{\theta})}{\partial \theta_j} = \sum\limits_{k=1}^{N_j}
    {\beta_k^{(j)}\langle X_0\otimes O \rangle_k}
\end{equation}

It is worth mentioning that the HT method has numerous variants \cite{guerreschi2017practical, ekert2002direct, li2017efficient, romero2018strategies}. The original HT methods employ the well-known Hadamard test to evaluate Eq.~\eqref{eq:grad_def_HT}. While no measurements are required on the other qubits except for the ancilla, this approach demands a significantly higher number of controlled gates. The version presented here \cite{guerreschi2017practical} also measures the other qubits, thus enabling more advanced measurement optimization techniques, as will be discussed in Section~\ref{sec:measurement_grouping}.

\subsubsection{Direct Hadamard Test (DHT)}
\label{sec:DHT}
DHT \cite{garcia2013swap, indirect_to_direct} is an alternative method to the traditional HT. DHT eliminates the need for an ancillary qubit and the associated extra CNOT gates, but it requires \(2^p\) distinct circuits to compute the same value when there are \(p\) CNOT gates involving the ancilla in the standard HT. DHT is illustrated in Fig.~\ref{fig:grad_methods}(c), with additional details provided in Appendix~\ref{sec:appendix_DHT}. DHT can be considered a generalization of the naive PSR for general PQCs, with PSR being a specialized method within the broader family of direct methods. For measuring gradients as in Eq.~\eqref{eq:grad_def_HT}, DHT incurs twice the number of distinct circuits needed, while reducing the number of CNOT gates in each circuit.

\subsection{Measurement Optimization}
\label{sec:measurement_grouping}

In parameterized quantum circuits, the goal is to calculate the expectation values of observables for a prepared quantum state, \(\langle O \rangle = \bra{\boldsymbol{\theta}} O \ket{\boldsymbol{\theta}}\), as in Eq.~\eqref{eq:f(theta)}. These observables \(O\) can be written as a linear combination of \(\mathcal{N}(O)\) Pauli words: 
\[
O = \sum_{l=1}^{\mathcal{N}(O)}\alpha_l P_l, \quad P_l \in \{I, X, Y, Z\}^{\otimes N}.
\]
In many variational quantum algorithms, the number of terms can be prohibitively large. For example, in VQE, large molecules can commonly induce \(\mathcal{N}(O) \sim N^4\) terms in the observable \cite{verteletskyi2020measurement, wecker2014gate, hastings2014improving, Jordan:1928wi, seeley2012bravyi, bravyi2002fermionic, measurement_grouping_yongshan}. Naively, we need \(\mathcal{N}(O)\) measurements, one for each \(P_l\), to evaluate \(\langle O \rangle = \sum_{l=1}^{\mathcal{N}(O)}\alpha_l \langle P_l \rangle\). Fortunately, measurement grouping techniques can significantly reduce this burden.

If two Pauli terms \(P_{l_1}\) and \(P_{l_2}\) commute, i.e., \([P_{l_1}, P_{l_2}] = 0\), they share a common eigenbasis \(\{\phi_r\}_r\), which simultaneously diagonalizes both:
\[
P_{l_1} = \sum_r \lambda_{l_1, r} \ketbra{\phi_r}, \quad P_{l_2} = \sum_r \lambda_{l_2, r} \ketbra{\phi_r}.
\]
The expectation values of these terms can then be expressed as:
\[
\langle P_{l_1} \rangle = \sum_r \lambda_{l_1,r} \norm{\braket{\phi_r}{\boldsymbol{\theta}}}^2, \quad \langle P_{l_2} \rangle = \sum_r \lambda_{l_2,r} \norm{\braket{\phi_r}{\boldsymbol{\theta}}}^2.
\]
By performing a single measurement in their shared eigenbasis \(\{\phi_r\}_r\), both expectation values can be obtained simultaneously.

The measurement grouping procedure is outlined in Algorithm \ref{alg:meas_grouping}. Given the observable Hamiltonian \(O = \sum_{l=1}^{\mathcal{N}(O)}\alpha_l P_l\), classical algorithms are used to partition the \(\mathcal{N}(O)\) terms into \(\mathcal{N}_{cm}(O)\) groups:
\[
O = \sum_{s=1}^{\mathcal{N}_{cm}(O)}\mathcal{G}_s, \quad \mathcal{G}_s = \sum_{t=1}^{T_s} \mathcal{C}_{s,t} P_{s,t}, \quad \sum_{s=1}^{\mathcal{N}_{cm}(O)} T_s = \mathcal{N}(O),
\]
where \(\mathcal{C}_{s,t}\) are the scalar coefficients for corresponding $P_{s,t}$ and all terms within a group \(\mathcal{G}_s\) commute:
\[
\forall s \in [\mathcal{N}_{cm}(O)]: \quad [P_{s,t}, P_{s,t'}] = 0, \quad \forall t, t' \in [T_s].
\]
By measuring all terms in each group simultaneously, the number of distinct measurements is reduced from \(\mathcal{N}(O)\) to \(\mathcal{N}_{cm}(O)\). For example in \cite{measurement_grouping_yongshan}, measurement grouping for VQE reduced the number of distinct circuit executions from \(O(N^4)\) to \(O(N^3)\).

\begin{algorithm}
\caption{Measurement Grouping}\label{alg:meas_grouping}
\begin{algorithmic}[1]
    \State Partition the \(\mathcal{N}(O)\) terms in \(O\) into \(\mathcal{N}_{cm}(O)\) groups.
    \State $\mu \gets 0$
    \For{each $\mathcal{G}_s$ of the \(\mathcal{N}_{cm}(O)\) groups}
        \State Apply unitary transformations that rotate the 
        \Statex \hspace{\algorithmicindent}eigenbasis \(\{\phi_r\}_r\) to the computational basis
        \Statex \hspace{\algorithmicindent}to the quantum states to be measured.
        \State Measure in the computational basis.
        \State $\mu_s \gets 0$
        \For{each $P_{s,t}$ of the $T_s$ operators in $\mathcal{G}_s$}
            \State Perform classical post-processing to recover the
            \Statex \hspace{\algorithmicindent}\hspace{\algorithmicindent}measurement expectation values $\langle P_{s,t} \rangle$.
            \State $\mu_s \pluseq C_{s,t} \cdot \langle P_{s,t} \rangle$
        \EndFor
        \State $\mu \pluseq \mu_s$
    \EndFor
    \State \Return $\mu$ as the estimate for the expectation value \(\langle O \rangle\).
\end{algorithmic}
\end{algorithm}

In this section, we have discussed the general principles of measurement grouping techniques. While different approaches \cite{grouping_1, grouping_efficient, grouping_graph_theory, grouping_unitary_partioning, measurement_grouping_yongshan} may vary in their partitioning algorithms and the implementation of pre-measurement unitary transformations, they all can be integrated into our gradient estimation framework. It is important to note that these different pre-measurement unitary transformation methods come with varying costs and complexities. The approach we discussed primarily uses full commutativity as the partitioning criterion, while other methods may use different criteria, such as pairwise qubit commutativity. The improvements provided by these techniques highlight the importance of selecting the right observables to measure, and our work provides the flexibility needed to make these optimal choices.

%% file: 3_gradient_circuits.tex
\section{Efficient Gradient Estimation}
\label{sec:efficient_grad}

In this section, we propose three techniques for more efficient gradient estimation. The first technique, based on the Hadamard test circuit, provides flexibility in selecting which Hermitian operator to measure when multiple operators are involved in the product observable, allowing for better measurement grouping. By applying this technique to both HT and DHT, we introduce two new gradient estimation methods. For higher-order derivatives, we propose an algorithm that potentially offers an exponential advantage over previous methods. This algorithm also benefits from some flexibility in choosing which operator to measure. Finally, we introduce a framework that combines various gradient estimation algorithms and, considering the trade-offs of each, automatically selects the optimal method for different scenarios. To the best of our knowledge, this is the first implementation of gradient estimation that can employ different algorithms for parameters within a single variational circuit.

\subsection{Flexible Hadamard Test}
\label{sec:flex_HT}

To introduce the algorithm, we first simplify some notations. We denote Hermitian operators:
\begin{align}
    \tilde{H}_j = U_{(j+1):n} H_j U_{(j+1):n}^\dagger
    \label{eq:def_herm_op}
\end{align}
Then the definition of gradient in Eq.~\eqref{eq:grad_def_HT} and Eq.~\eqref{eq:grad_def_com} can be simplified as:

\begin{align}
    \pdv{f(\boldsymbol{\theta})}{\theta_j} = -\Im{
    \bra{\boldsymbol{\theta}} \tilde{H}_j O \ket{\boldsymbol{\theta}}}
    \label{eq:simple_grad_def} = \frac{i}{2} \bra{\boldsymbol{\theta}} [\tilde{H}_j, O] \ket{\boldsymbol{\theta}} 
\end{align}
Note that the original Hadamard test circuit only provides a method to evaluate expressions such as \(\Im{\bra{\boldsymbol{\theta}} U \ket{\boldsymbol{\theta}}}\), where \(U\) is a unitary gate. Previous works \cite{ekert2002direct, li2017efficient, romero2018strategies} compute gradients by directly substituting such \(U\) with \(\tilde{H}_j O\). In contrast, \cite{guerreschi2017practical} discovered a more efficient circuit to evaluate Eq.~\eqref{eq:simple_grad_def} by measuring \(O\) directly instead of implementing a controlled-\(O\) gate, thereby enabling the use of measurement optimization techniques. Our flexible HT algorithm builds on this approach by allowing flexibility in choosing which operator to measure, rather than being restricted to measuring \(O\). This flexibility facilitates better measurement grouping and further optimizes the gradient estimation process.

We restate the problem to which our algorithm applies in a more general setting. This generalized version of the algorithm can address problems beyond gradient estimation in Eq.~\ref{eq:simple_grad_def}, which we will also discuss in the later sections of the paper. In this general setting, we do not differentiate the observable \(O\) and the generators \(\{\tilde{H}_j\}_j\), since they are all Hermitian operators mathematically, which are now collectively denoted by \(\{\hat{H}_m\}_m\).

\begin{definition}[Expectation Value of an \(m\)-Term Product]
\label{def:m-term_product_exp_value}
Given \(m\) Hermitian operators \(\hat{H}_1, \hat{H}_2, \cdots, \hat{H}_m\) and an input quantum state \(\ket{\boldsymbol{\theta}}\), the expectation value for measuring their product operator is defined as 
\[
\Bigg \langle \prod_{i=1}^{m} \hat{H}_i \Bigg \rangle_{\ket{\boldsymbol{\theta}}} = \bra{\boldsymbol{\theta}} \hat{H}_1 \hat{H}_2 \cdots \hat{H}_m \ket{\boldsymbol{\theta}}
\]
\end{definition}

In the following section, we present an algorithm for estimating the imaginary part of \(\langle \prod_{i=1}^{m} \hat{H}_i \rangle_{\ket{\boldsymbol{\theta}}}\). Note that minor modifications can be made to the algorithm to estimate the real part. Estimating the first-order gradient of a parameterized circuit is a special case when \(m=2\). The more general \(m\)-term cases appear in higher-order derivative estimation \cite{harrow2021low}, so we state the general case here to ensure the algorithm also applies to higher-order derivative estimation in Sec.~\ref{sec:k-fold-HT}.

Our constructed circuit allows for an arbitrary choice of these \(\hat{H}_i\) to be the measured observable. This technique improves over measuring a fixed observable by enabling the selection of the \(\hat{H}_i\) that offers the most efficiency gain after measurement optimization. For clarity, we assume all other operators that are not measured can be easily implemented as quantum gates. Otherwise, one can follow Eq.~\eqref{eq:hj_decomposition} for decomposition, as explained in detail in Appendix~\ref{sec:appendix_flex_HT_example}. The algorithm is stated in the following:

\begin{algorithm}
\caption{Estimate \(\Im{\langle \prod_{i=1}^{m} \hat{H}_i \rangle_{\ket{\boldsymbol{\theta}}}}\) via $\hat{H}_i$ Meas.}\label{alg:pick_ht}
\begin{algorithmic}[1]
    \State Initialize Reg. $A$ to $\frac{1}{\sqrt{2}}\left( \ket{0} - i\ket{1} \right)$; Reg. $B$ to $\ket{\boldsymbol{\theta}}_B$.
    \State Controlled on Reg. $A$ being 0, we apply $\hat{H}_1 \cdots \hat{H}_{i-1}$ to Reg. $B$.
    \State Controlled on Reg. $A$ being 1, we apply $\hat{H}_{m} \cdots \hat{H}_{i+1}$ to Reg. $B$.
    \State Measure the observable $X \otimes \hat{H}_i$ on Reg. $A$ and $B$, with measurement optimization techniques in Algorithm.~\ref{alg:meas_grouping}.
\end{algorithmic}
\end{algorithm}

\begin{figure}[h]
    \captionsetup{justification=raggedright}
    \centering
    \includegraphics[width=\linewidth]{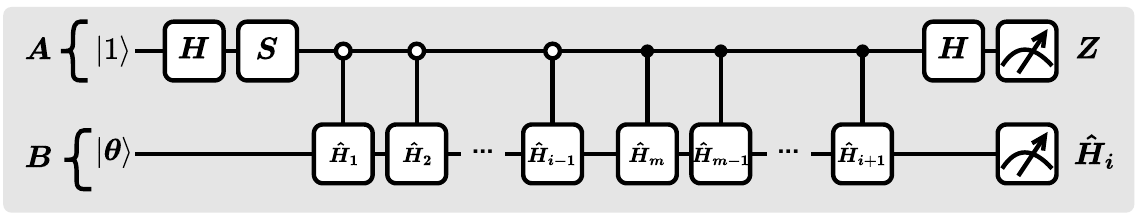}
    \caption{Flexible Hadamard Test}
    \label{fig:flexible_HT}
\end{figure}

Detailed explanations and proofs are provided in Appendix~\ref{sec:appendix_flex_HT}. Next we will show how to apply Algorithm.~\ref{alg:pick_ht} to estimating the first-order gradient of parameterized quantum circuits and elaborate the generated algorithms.

\subsubsection{Reversed Hadamard Test Gradient Method (RHT)}
\label{sec:RHT}

In the case of estimating the first-order gradient of variational circuits, we only have the gate's generator \(\tilde{H}_j\) and circuit observable \(O\) in the expression as in Eq.~\eqref{eq:simple_grad_def}. Our technique can be seen as exchanging the roles of gate generators and observables. Compared to \cite{guerreschi2017practical}, our method can estimate Eq.~\eqref{eq:simple_grad_def} by implementing gates using \(O\) and measuring \(\tilde{H}_j\) instead. We call this approach the Reversed Hadamard Test (RHT). One can choose to either measure \(O\), as in \cite{guerreschi2017practical}, or measure \(\tilde{H}_j\) by applying Algorithm~\ref{alg:pick_ht}, thus leveraging the flexibility to exchange the roles of these two terms for gradient estimation.

Notably, \cite{anastasiou2023really} explores similar relationships within a specific quantum ansatz class, focusing on gradients in the rear gates. In contrast, our work addresses a more general setting, demonstrating the broader applicability of the reversed method.

More concretely, similar to HT in Section.~\ref{sec:HT}, we denote the decomposition of the observable $O$ as:

\begin{equation}
    O = \sum\limits_{l=1}^{M}\alpha_lP_l,
    \label{eq:o_decomposition}
\end{equation}
Then for each Pauli word $P_l$ in the decomposition, the method uses a reversed Hadamard test circuit as illustrated in Fig.~\ref{fig:grad_methods}(d) to evaluate:
\begin{align}
    \langle X_0 \otimes H_j \rangle_l = -\Im{
    \bra{\theta} \tilde{H}_j P_l \ket{\theta}}
\end{align}
where $X_0 \otimes H_j$ is the observable. We need a total of $M$ such circuits executed to obtain the gradient values.

\begin{equation}
    \frac{\partial f(\boldsymbol{\theta})}{\partial \theta_j} = \sum\limits_{l=1}^{M}
    {\alpha_l\langle X_0 \otimes H_j \rangle_l}.
\end{equation}

The proof of correctness can be found in Appendix~\ref{sec:appendix_RHT}.

Compared to HT, a key quantity of interest is the distinct circuit execution count, which approximates the actual run time on real hardware. From Section~\ref{sec:measurement_grouping}, we know that commuting terms in the observable can be grouped to reduce the execution count. Intuitively, in the context of HT and RHT, with the help of measurement grouping, when the terms in the observable \(O\) are more groupable, HT will require fewer distinct circuit executions to evaluate the gradient. Conversely, when the terms in the generator \(H_j\) are more groupable, DHT would be the better choice. This distinction is illustrated in Table.~\ref{tab:models}, with examples provided in the caption.

Such groupable Hamiltonians are widely used as generators in variational algorithms such as QAOA and VQE. For instance, in QAOA for MaxCut, all terms in the cost layer generator and mixing layer generator commute. Similarly, in VQE with UCCSD, all terms in the generator of the single-excitation and double-excitation layers commute. These applications, where it is beneficial to group the generators, are ideal use cases for RHT. This observation informs the experimental design in Section~\ref{sec:power_reversed}, where we numerically study the effectiveness of the reversed methods.

\subsubsection{Reversed Direct Hadamard Test Gradient Method (RDHT)}
\label{sec:RDHT}

Similar techniques can be applied for DHT so that the roles of gate generators and observables will be exchanged, as illustrated in Fig.~\ref{fig:grad_methods}(e). More details can be found in Appendix.~\ref{sec:appendix_RDHT}.

\begin{figure*}[t]
    \captionsetup{justification=raggedright}
    \includegraphics[width=\linewidth]{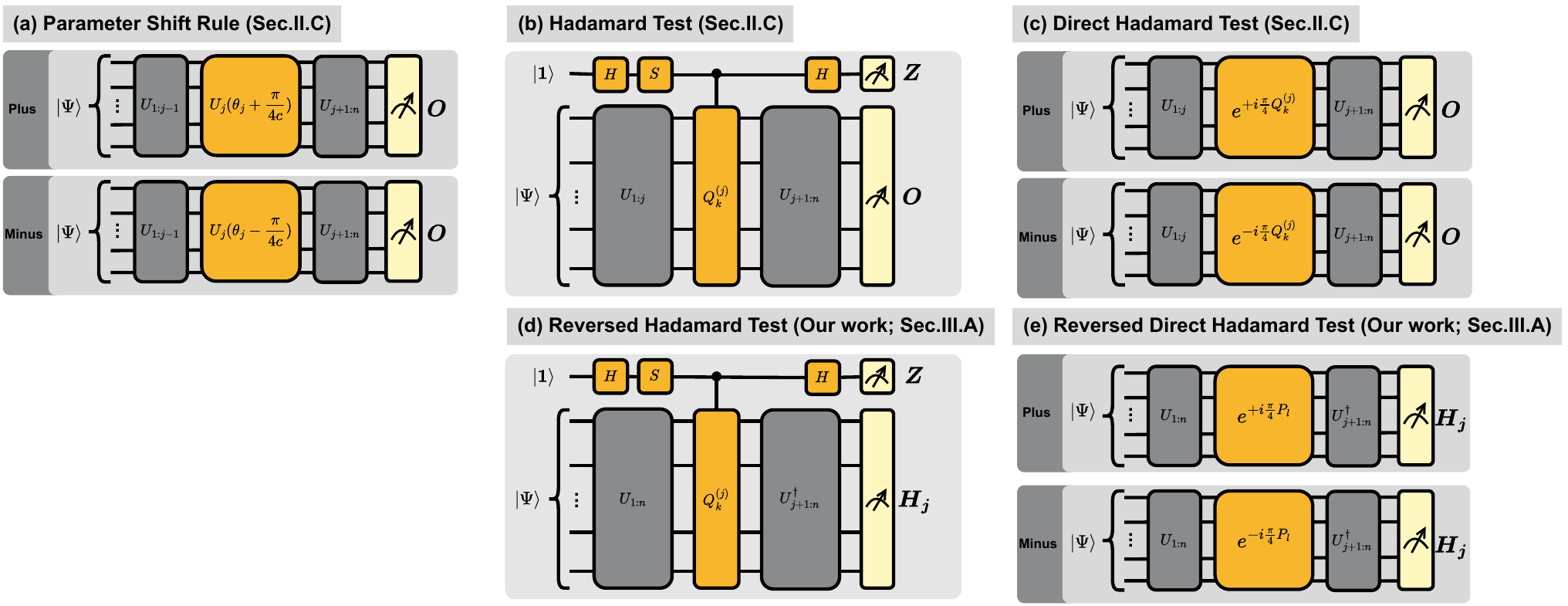}
    \caption{Circuit implementation of gradient methods. (a) PSR changes the parameter values of the original circuit for gradient evaluation (b) HT requires one ancillary qubit and controlled multi-qubit gates (c) DHT replaces the indirect measurements in HT but doubles the circuit execution needed (d) RHT invert the roles of generator $H_j$ and observable $O$ (e) RDHT combines DHT and RHT. }
    \label{fig:grad_methods}
\end{figure*}

\subsection{Higher-order Derivative}
\label{sec:k-fold-HT}

Besides first-order derivatives, higher-order derivatives such as the Hessian or the Fubini-Study metric (Fisher information metric) \cite{facchi2010classical, meyer2021fisher, spehner2017geometric}, are also important for parameterized quantum circuits. The \(k^{\text{th}}\)-order partial derivative of \(f(\boldsymbol{\theta})\), as defined in Eq.~\eqref{eq:f(theta)}, is given by:

\[
\frac{\partial^k f}{\partial \theta_{j_1} \partial \theta_{j_2} \cdots \partial \theta_{j_k}} (\boldsymbol{\theta}),
\]

where \(j_1, j_2, \ldots, j_k\) are indices arbitrarily selected from \(\{1, 2, \ldots, n\}\). By the symmetry of mixed partial derivatives \cite{harrow2021low}, the order of differentiation does not affect the result:

\[
 \frac{\partial^k f}{\partial \theta_{j_{\sigma(1)}} \partial \theta_{j_{\sigma(2)}} \cdots \partial \theta_{j_{\sigma(k)}}} (\boldsymbol{\theta}) = \frac{\partial^k f}{\partial \theta_{j_1} \partial \theta_{j_2} \cdots \partial \theta_{j_k}} (\boldsymbol{\theta})
\]

for any permutation \(\sigma\) of \(\{1, 2, \ldots, k\}\). Therefore, without loss of generality, we can assume \(1 \leq j_1 \leq j_2 \leq \cdots \leq j_k \leq n\). Thus, we define the \(k^{\text{th}}\)-order partial derivative as follows:

\begin{definition}[$k^{\text{th}}$-order Partial Derivative]
\label{def:k-th_order_derivative}
For the target function $f(\boldsymbol{\theta})$ defined in Eq.~\eqref{eq:f(theta)}, given any $k$ indices \(1 \leq j_1 \leq j_2 \leq \cdots \leq j_k \leq n\), the \(k^{\text{th}}\)-order partial derivative is defined as
\begin{align}
    &\frac{\partial^k f}{\partial \theta_{j_1} \partial \theta_{j_2} \cdots \partial \theta_{j_k}} (\boldsymbol{\theta}) \nonumber \\
    =\quad & \left(\frac{i}{2}\right)^k \bra{\boldsymbol{\theta}} [\tilde{H}_{j_1}, [\tilde{H}_{j_2}, [\cdots, [\tilde{H}_{j_k}, O]]]] \ket{\boldsymbol{\theta}}
    \label{eq:k-th_derivative}
\end{align}
\end{definition}

Therefore, when referring to the \(k^{\text{th}}\)-order derivative, we will assume the indices are ordered as such, since the permutation of the indices does not affect the outcome. Interested readers can find the proof of Eq.~\eqref{eq:k-th_derivative} in Appendix.~\ref{sec:appendix_k_th_order_value}. The nested commutator in Eq.~\eqref{eq:k-th_derivative}, upon expansion, generally results in \(2^k\) distinct terms, making evaluation challenging. In PSR \cite{vanilla_PSR, crooks_PSR, fourtermPSR, stochasticPSR, generalPSR, mitarai2018quantum, mari2021estimating}, this nested commutator can be simplified when the generators \(\{ H_{j_t} \}_{t=1}^k\) of the parameterized circuits have only two distinct eigenvalues. However, even in this special case, evaluating Eq.~\eqref{eq:k-th_derivative} requires \(2^k\) distinct circuit executions. Furthermore, PSR is not applicable if any of the \(\{ H_{j_t} \}_{t=1}^k\) has more than two distinct eigenvalues. Similar to the first-order derivative case, hadamard test based methods can handle the more complex cases. The existing method \cite{harrow2021low} runs a distinct circuit for each term in the expansion of the nested commutator, which in total results in $2^k$ distinct circuit execution to evaluate Eq.~\eqref{eq:k-th_derivative}. Note that each of the $2^k$ terms can employ Algorithm.~\ref{alg:pick_ht} to reduce cost and speed up the estimation. Another method is to to use DHT, however for a HT circuit that contains $p$ controlled-gates, DHT needs to run $2^p$ distinct circuits to estimate it \cite{indirect_to_direct}. So to evaluate Eq.~\eqref{eq:k-th_derivative}, DHT needs to evaluate $2^{k-1}\cdot2^k$ distinct circuits. However, for DHT we have developed a more efficient algorithm to evaluate the $k^{th}$-order derivative which only uses $2^k$ distinct circuits. Interested readers can find the algorithm in Appendix.\ref{sec:appendix_DHT_for_k}.

Here we present an algorithm based on the Hadamard test to evaluate Eq.~\eqref{eq:k-th_derivative}, as defined in the \(k^{\text{th}}\)-order partial derivative in Definition \ref{def:k-th_order_derivative}, using only one quantum circuit. The algorithm utilizes \(k\) ancillary qubits. In practical applications, such as when evaluating the second-order derivative, this typically requires only two additional qubits.

\begin{algorithm}
\caption{Est. $\left(\frac{i}{2}\right)^k \bra{\boldsymbol{\theta}} [\tilde{H}_{j_1}, [\tilde{H}_{j_2}, [\cdots, [\tilde{H}_{j_k}, O]]]] \ket{\boldsymbol{\theta}}$}
\label{alg:higher_ht}
\begin{algorithmic}[1]
    \State Initialize register $A$ to $\left( \frac{1}{\sqrt{2}} \left( \ket{0} - i\ket{1} \right) \right)^{\otimes k}$.
    \State Initialize register $B$ to $\ket{\boldsymbol{\theta}}_B$.
    \State For $i = 1$ to $k$, apply the controlled operation $\tilde{H}_{j_i}$ on register $B$, controlled by the $i$-th qubit of register $A$ being $\ket{1}$.
    \State Measure the observable $X^{\otimes k} \otimes O$ on the combined system of registers $A$ and $B$, with measurement optimization techniques in Algorithm.~\ref{alg:meas_grouping}.
\end{algorithmic}
\end{algorithm}

\begin{figure}[h]
    \captionsetup{justification=raggedright}
    \centering
    \includegraphics[width=\linewidth]{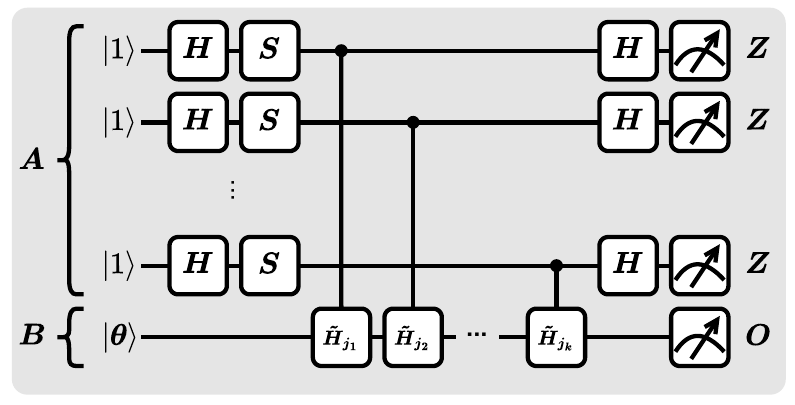}
    \caption{$k^{th}$-order Derivative Estimation}
    \label{fig:k-th_order_gradient_algo}
\end{figure}

We illustrate Algorithm.\ref{alg:higher_ht} with Fig.\ref{fig:k-th_order_gradient_algo}, where if \(\{ H_{j_t} \}_{t=1}^k\) are difficult to implement as gates, one can follow similar decomposition techniques in Appendix.~\ref{eq:appendix_flex_HT_decomposition}. This nice generalization of the single qubit Hadamard test circuit turns out, surprisingly, to be measuring the nested commutator in Eq.~\eqref{eq:k-th_derivative}. The proof of correctness can be found in the Appendix.\ref{sec:appendix_k_th_order_alg}. As summarized in Table.~\ref{tab:k_methods}, our algorithm saves exponential distinct circuit executions and sample complexity, which makes it the ideal algorithm for evaluating higher-order derivatives.

\begin{table*}[]

    \captionsetup{justification=raggedright}
    \centering

    \resizebox{\textwidth}{!}{
    \begin{tabular}{p{0.28\textwidth}p{0.18\textwidth}p{0.15\textwidth}p{0.15\textwidth}p{0.24\textwidth}}
    \hline\hline
        \textbf{Method} &  \textbf{Number of $\quad\quad\quad\quad$ Distinct Circuits} & \textbf{Qubits} & \textbf{Depth} &\textbf{Algorithmic limitations} \\ \toprule
    Parameter shift rule (PSR) \cite{mitarai2018quantum} & $2^k$ & $N$ & $n$ & Restricted to simple $\{H_j\}_j$  \\\hline
    Hadamard Test (HT) \cite{guerreschi2017practical}& $2^{k-1}$ & $N+1$ & $n+k$ & General $\{H_j\}_j$ allowed \\\hline
    Direct Hadamard Test (DHT) \cite{indirect_to_direct}& $2^k$ & $N$ & $n+k$ &General $\{H_j\}_j$ allowed\\\hline
    
    $k$-fold Hadamard Test (Our work) & 1  & $N+k$ & $n+k$ & General $\{H_j\}_j$ allowed\\
    \hline\hline
    \end{tabular}
}
    \vspace{5pt}
    \caption{Quantum methods for estimating \(k^{\text{th}}\)-order partial derivatives as defined in Def.~\ref{def:k-th_order_derivative}. Here, \(N\) represents the number of qubits in the given parameterized quantum circuit. $n$ is the number of parameterized gates in the original circuit as illustrated in Fig.~\ref{fig:qc_ansatz}. Note that for DHT, the original method used in \cite{indirect_to_direct} required \(2^{k-1} \cdot 2^k\) gradient circuits. However, we derived a more efficient algorithm in Appendix \ref{sec:appendix_DHT_for_k} that uses only \(2^k\) circuits.
    }
    \label{tab:k_methods}
\end{table*}

Note that although Algorithm.\ref{alg:higher_ht} does not enjoy the flexibility of choosing whichever Hamiltonian to measure like Algorithm.\ref{alg:pick_ht} does, it still allows measurement optimization by picking one of the inner most Hamiltonians, i.e. $\tilde{H}_{j_k}$ and $O$, as the measurement. This observation is useful in practical implementation for higher-order derivative estimation but will not be the main focus of our work.

\subsection{Quantum Automatic Differentiation Software}
\label{sec:QAD}

\begin{figure}[h]
    \centering
    \includegraphics[width=\linewidth]{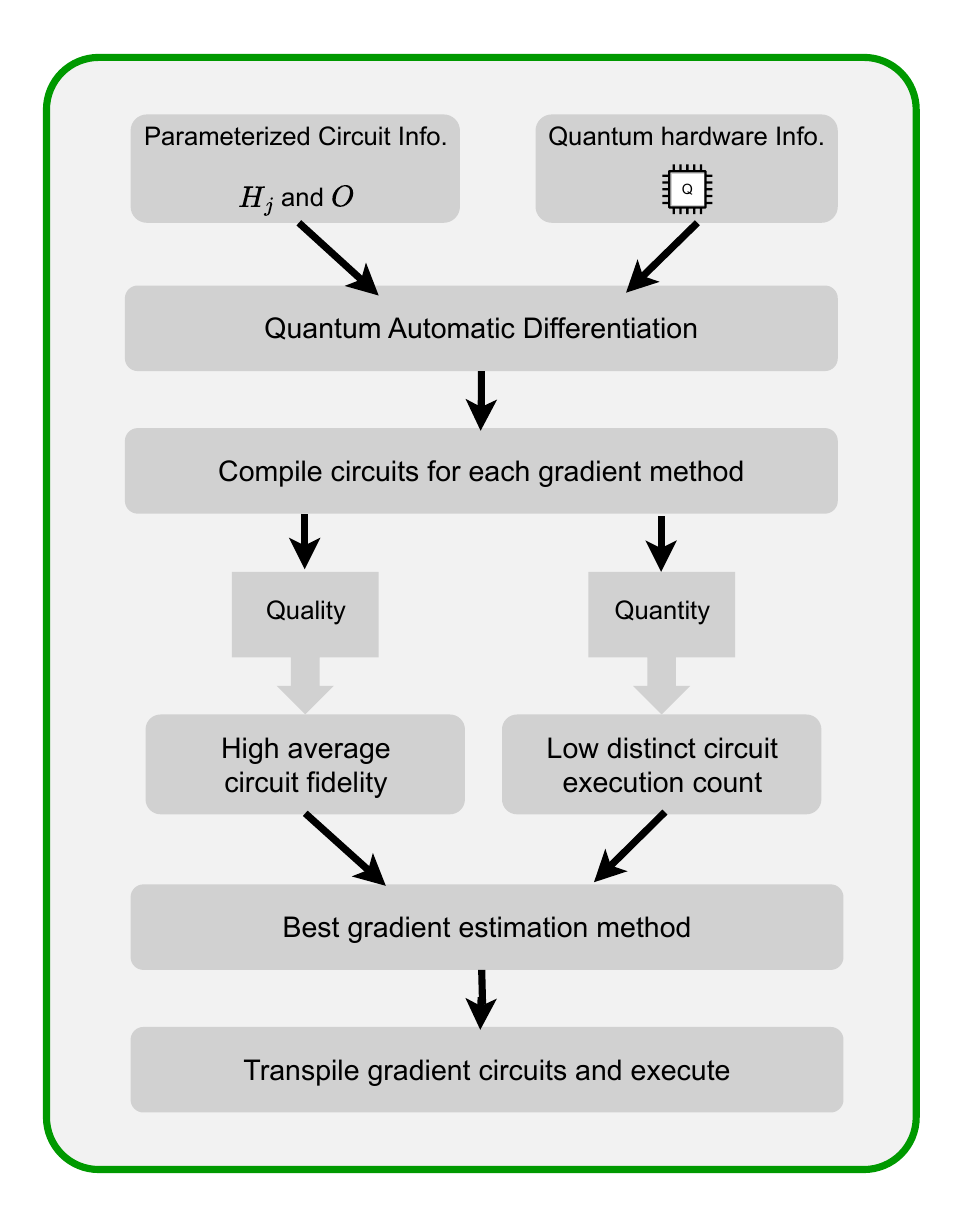}
    \caption{Overview of the QAD method}
    \label{fig:flow}
\end{figure}

As discussed in previous sections, different gradient estimation methods result in different gradient circuits that need to be executed. These methods vary in terms of the \textit{number of distinct circuits} and \textit{circuit depth}, as illustrated in Table \ref{tab:models}. In this section, we introduce the Quantum Automatic Differentiation (QAD) method, a comprehensive framework that integrates various gradient techniques, including PSR, HT, DHT, RHT, and RDHT, for enhanced efficiency in computing PQC gradients. QAD's distinctive approach involves assessing each parameter in a PQC on an individual basis, considering the unique attributes of each \(H_j\) and \(O\). As a result, QAD dynamically assigns the optimal gradient method for each parameter, which allows QAD to leverage the strengths of each gradient method, significantly enhancing overall performance compared to the utilization of a singular method for all parameters. While the flexibility of measurements in Algorithm \ref{alg:higher_ht} enables the definition of QAD for higher-order gradient estimation, our focus in the rest of the paper is on first-order gradient estimation. This focus will also be reflected in the numerical experiments presented later.

\begin{table*}[]
    \captionsetup{justification=raggedright}
    \centering

    \resizebox{\textwidth}{!}{
    \begin{tabular}{p{0.368\textwidth}p{0.27\textwidth}p{0.16\textwidth}p{0.18\textwidth}}
    \hline\hline
        \textbf{Method} &  \textbf{Number of Distinct Circuits} & \textbf{Qubits} & \textbf{Depth}  \\ \toprule
    Parameter shift rule (PSR) \cite{vanilla_PSR, crooks_PSR, fourtermPSR, stochasticPSR, generalPSR, mitarai2018quantum, mari2021estimating} & $2 \times \mathcal{N}(H)\mathcal{N}_{cm}(O)$ & $N$ & $n$\\\hline
    Hadamard Test (HT) \cite{guerreschi2017practical, ekert2002direct, li2017efficient, romero2018strategies}& $\mathcal{N}(H)\mathcal{N}_{cm}(O)$ & $N+1$& $n+1$ \\\hline
    Direct Hadamard Test (DHT) \cite{garcia2013swap, indirect_to_direct}& $2 \times \mathcal{N}(H)\mathcal{N}_{cm}(O)$ & $N$ & $n+1$ \\\hline
    Reversed HT (Our work) & $\mathcal{N}_{cm}(H)\mathcal{N}(O)$ & $N+1$& $2n-j+1$\\\hline
    Reversed DHT (Our work) & $2 \times \mathcal{N}_{cm}(H)\mathcal{N}(O)$ & $N$ & $2n-j+1$\\\hline
    QAD Method (Our work) & $\min$ per chosen metric & method-dependent & $\min$ per chosen metric\\
    \hline\hline
    \end{tabular}
}
    \vspace{5pt}
    \caption{Quantum first-order gradient methods with measurement optimization for the $j^{th}$ parameter. We denote the generator $H_j$ of the $j^{th}$ gate as $H$ for simple notations. As defined in Sec.~\ref{sec:measurement_grouping}, $\mathcal{N}(\cdot)$ denotes the number of Pauli terms in the linear combination; $\mathcal{N}_{cm}(\cdot)$ denotes the number of disjoint subsets after grouping by commutativity. For instance, $H = 1 * Z_0Z_1 + 1*X_0X_1 + 1*Z_0X_1$, $\mathcal{N}(H) = 3$ because it has three terms and $\mathcal{N}_{cm}(H) = 2$ because $H = \left(1 * Z_0Z_1 + 1*X_0X_1\right) + \left(1*Z_0X_1\right)$, where terms in each subset commute. $N$ is the number of qubits and $n$ is the number of parameterized gates. QAD can either optimize for fewer circuits or higher circuit fidelity, and for such chosen metric, QAD would pick the best gradient method for each individual parameter.}
    \label{tab:models}
\end{table*}

QAD, as illustrated in Fig.~\ref{fig:flow}, for each individual parameter, (1) it first identifies the feasible gradient methods; (2) For all feasible gradient methods, QAD constructs and compiles quantum circuits that need to be executed for gradient calculation corresponding to that parameter; (3) Utilizing either approximated circuit fidelity or distinct circuit execution count as the metric function, QAD computes a scalar cost value for each gradient method; (4) Based on these cost evaluations, QAD strategically assigns the most effective gradient method for each PQC parameter; (5) Employing the selected methods, QAD then constructs the gradient circuits, which are integral to the optimization process. 

Further experimental demonstration on QAD's capabilities and advantages is provided in Section~\ref{sec:evaluation}, illustrating its potential to revolutionize gradient calculation in variational quantum algorithms.

Note that although QAD requires a more complex compilation process before executing the quantum circuits on real hardware, in the VQA setting, this compilation is only needed once before the training begins. Throughout all iterations, the \(H_j\) and \(O\) remain unchanged, eliminating any additional compilation overhead after the initial setup.

%% file: 4_evaluation.tex
\section{Evaluation}
\label{sec:evaluation}

To better understand the differences and trade-offs between the individual methods including HT, DHT, RHT and RDHT, and to demonstrate the effectiveness of the unified QAD algorithm, we design a series of numerical experiments. First, we investigate the Quantum-Assisted Quantum Compiling problem, showcasing how the flexibility provided by Algorithm \ref{alg:pick_ht} combined with measurement grouping techniques can lead to significant improvements for VQAs. Next, we study a Quantum Neural Network to illustrate how QAD can handle networks composed of various gates, each benefiting from different gradient methods. QAD treats each parameter individually, resulting in substantial improvements. In the following sections, we first present the benchmark experiments, then their results and interpretation.

\remark{Note that we also conducted experiments on QAOA for MaxCut to compare the advantages provided by the direct methods (DHT and RDHT). Overall, our findings indicate that direct methods can produce higher quality gradient estimation circuits, albeit at the cost of doubling the number of circuits required. A more detailed analysis can be found in Appendix \ref{sec:appendix_direct}.}

\subsection{Experimental Setup}
We conducted all experiments using PennyLane framework \cite{pennylane} and leverage the Qiskit \cite{Qiskit} backend to execute quantum circuits. Instead of using real quantum hardwares, which have very limited access, long queueing time and are not feasible for variational experiments of our scale, we use backend simulators that use a noise model built from real hardware system snapshots. In particular, we choose the \textit{ibm\_lagos} and \textit{ibm\_cairo} device, which contain 7 and 27 qubits respectively. The snapshots contain crucial information about the quantum system, including device topology, basis gates, and error rates. The noise model built from real device calibration data is more realistic than theoretical noise models such as the depolarizing noise. For each benchmark task, our baseline gradient methods are PSR, HT, DHT, RHT, RDHT and QAD. Each experiment is repeated 5 times with different random seeds to provide reliable conclusions.

\subsection{Quantum Assisted Quantum Compiling}
\label{sec:power_reversed}

\textbf{Benchmarks:} As mentioned in Section \ref{sec:flex_HT}, with the aid of measurement grouping, reversed methods can offer advantages when the generator \(H_j\) is more groupable than the observable \(O\). This is quantified in terms of the distinct circuit execution count. In NISQ implementations, unique circuit configurations tends to be more costly than generating samples with configured circuits \cite{generalPSR}. So distinct circuit execution count often reflects the actual runtime required on real quantum hardware and is crucial for optimization tasks. In this experiment, we will investigate how reversed methods can accelerate the convergence of variational quantum algorithms (VQAs). Specifically, we will examine the variational algorithm known as Quantum Assisted Quantum Compiling (QAQC) \cite{khatri2019quantum, sharma2020noise}.

\begin{figure}[]
    \captionsetup{justification=raggedright}
    \centering
    \includegraphics[width=0.8\linewidth]{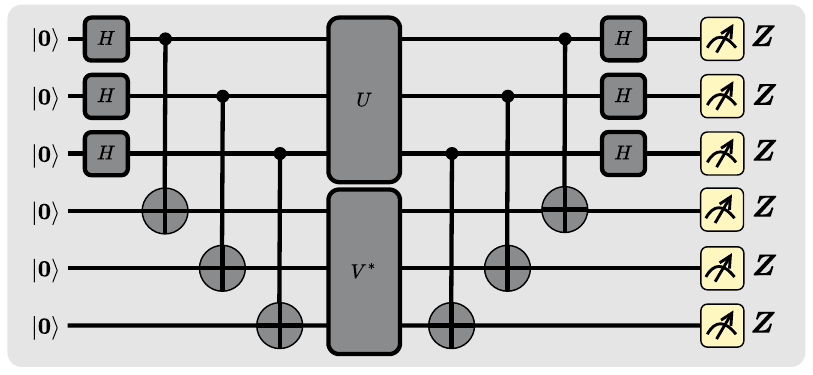}
    \caption{Hilbert-Schmidt Test}
    \label{fig:HST}
\end{figure}

To briefly introduce the problem setting: one is given a (analog) quantum computer which can evolve quantum systems via a time-dependent Hamiltonian $H$,
\begin{equation}
    H(t) = \sum_i \theta_i(t)H_i
\end{equation}
and a given unitary $U$ as the target (in a circuit-level representation), one needs to compute a pulse sequence $\theta_i(t)$ so that the (analog) quantum computer approximately performs $U$. QAQC discretizes the continuous pulse sequences into $k$ layers and solves the problem with a variational quantum algorithm. Specifically, QAQC aims to learn parameters $\{\theta_{k, i}\}_{k,i}$ so that the proposed circuit $V$, defined in Eq.\eqref{eq:QAQC_V}, is close to $U$.

\begin{equation}
    V = \prod_k \prod_i e^{-i\theta_{k, i}H_i}
    \label{eq:QAQC_V}
\end{equation}
In \cite{khatri2019quantum}, the similarity of $U$ and $V$, as the cost function, is measured by the Hilbert-Schmidt Test
\begin{align}
    C_{\text{HST}}(U,V)&= 1-\frac{1}{d^2}\left|\langle V,U\rangle \right |^2 \nonumber \\
    &=1-\frac{1}{d^2}| Tr (V^\dagger U)|^2
\end{align}
which can be evaluated through the quantum circuit in Fig.~\ref{fig:HST}

\begin{figure*}[]
    \captionsetup{justification=raggedright}
    \centering
    \includegraphics[width=\textwidth]{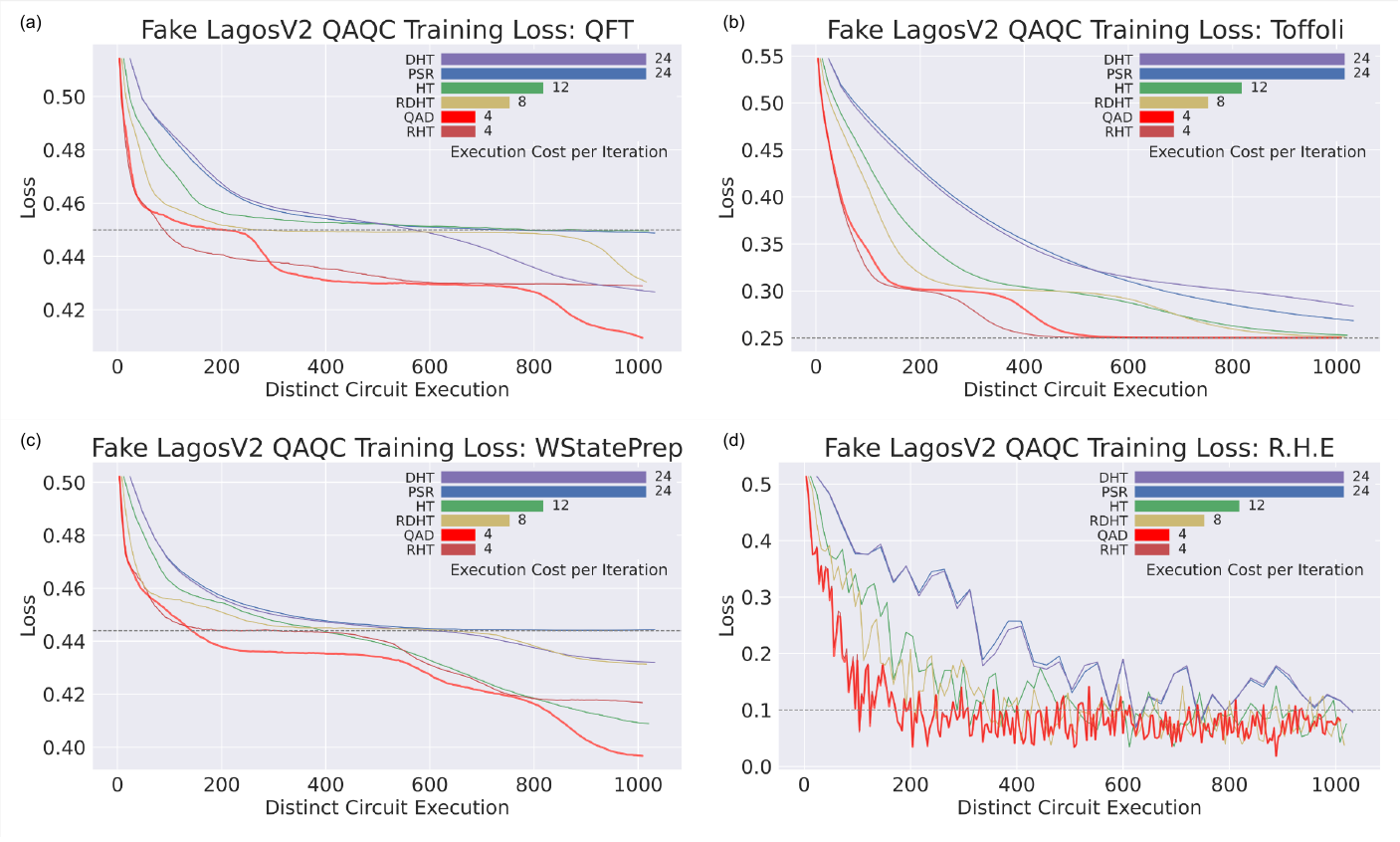}
    \caption{
    \textbf{(a)} the training loss curve for QFT \textbf{(b)} the training loss curve for Toffoli gate \textbf{(c)} the training loss curve for W-state preparation \textbf{(c)} the training loss curve for $U$ in Eq.~\eqref{eq:QAQC_efficient}. The mini bar plots in the top right corners are the cost per iteration for each gradient method. Specifically, the cost here is the distinct circuit execution count, which depends on the generator $H_j$ and observable $O$, and is not dependent on the target $U$ we want to learn.
    }
    \label{fig:reversed_power}
\end{figure*}

\begin{figure}[]
    \captionsetup{justification=raggedright}
    \centering
    \includegraphics[width=0.85\linewidth]{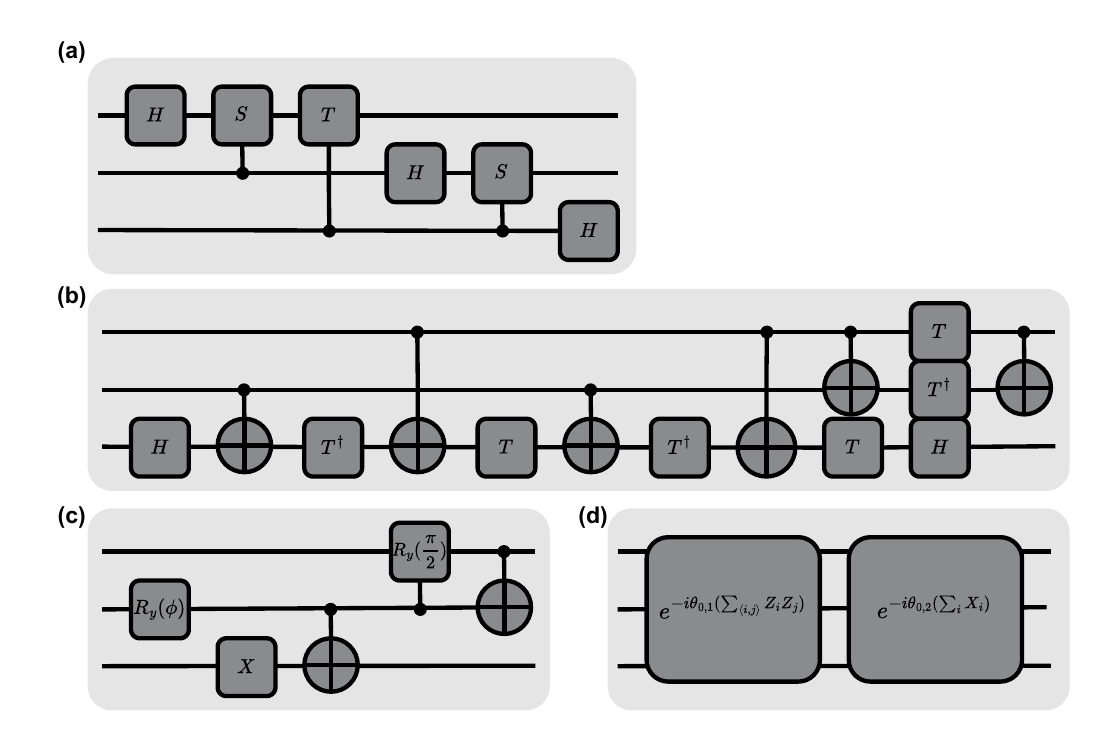}
    \caption{QAQC Benchmark Circuits: (a) Quantum Fourier transform (b) Toffoli gate (c) W-state preparation, where $\cos(\phi) = \frac{1}{\sqrt{3}}$ (d) Hardware Hamiltonian-inspired circuit}
    \label{fig:qaqc_circuits}
\end{figure}

Thus QAQC becomes a variational quantum algorithm and gradient methods come into play. Inspired by the hardware Hamiltonians of recent large superconducting quantum computers \cite{kim2023evidence} and neutral atom quantum computers \cite{wurtz2023aquila}, where in the decomposition of Eq.~\eqref{eq:hj_decomposition}, $N_j$ scales as $O(N)$, in our experiment, we simulate the toy example where the hardware Hamiltonian $H$ is a transverse-field Ising model \cite{pfeuty1970one}

\begin{equation}
    H(t) = \theta_1(t) \sum_{\langle i, j \rangle} Z_iZ_j + \theta_2(t) \sum_i X_i
\end{equation}
where $\langle i, j \rangle$ denotes the pairs of nearest neighbours among qubits. With first-order trotterization, the circuit $V$ is:
\begin{equation}
    V = \prod_k e^{-i\theta_{k,1}(\sum_{\langle i, j \rangle} Z_iZ_j)} \cdot e^{-i\theta_{k,2}(\sum_i X_i)}
\end{equation}
In this experiment we set $k=2$ and number of qubits in $H$ to be 3. For the benchmarks $U$, as illustrated in Fig.~\ref{fig:qaqc_circuits} we adopt the three hard circuits in \cite{sharma2020noise}, namely (1) Quantum Fourier Transform (2) Toffoli gate (3) W-state Preparation, and (4) we add one more relatively easier $U$:
\begin{equation}
    U = e^{-i\theta_{0,1}(\sum_{\langle i, j \rangle} Z_iZ_j)} \cdot e^{-i\theta_{0,2}(\sum_i X_i)}
    \label{eq:QAQC_efficient}
\end{equation}
where $\theta_{0,1}$ and $\theta_{0,2}$ are randomly chosen fixed numbers.

Note that adding benchmark (4) helps demonstrating the effectiveness of QAQC itself. but the advantages of reversed methods exist no matter what the target $U$ we want to learn, as long as the hardware Hamiltonian $H_j$ is more groupable.

\textbf{Baselines:} In this experiment, we compare all the gradient methods, including an additional classical baseline. For this baseline, we use classical backpropagation for gradient estimation, and the final result after convergence is compared with those obtained from the four quantum gradient methods. It is important to note that classical backpropagation is feasible only for small-scale circuits and does not scale to larger quantum problems. Therefore, we include this resource-intensive approach solely for comparison in this experiment, acknowledging that it is impractical for larger tasks. In this context, QAD consistently selects RHT for all parameters due to its significant advantage in reducing the circuit execution count per iteration.

\textbf{Results:} The experiment is simulated using the 7-qubit \textit{ibm\_lagos} machine with hardware-calibrated noises. As shown in Fig.~\ref{fig:reversed_power}, the reversed methods outperforms the non-reversed ones by $3\times$ in terms of circuit execution cost per iteration. With a more detailed theoretical derivation, reversed methods can achieve an $O(N)$ times speed up in QAQC for the hardware options we discussed above. Note that the simulation is done for transverse-field Ising model, but the algorithm could readily be deployed on real hardwares for the compiling tasks.
As a side note, we observe that QAQC with quantum gradient matches the backpropagation performances, and surprisingly, for some $U$, classical backpropagation is stuck at local minimums while QAQC can get out of those local minimums (the curves get flat at some point but they continue to go down with more iterations). \cite{sharma2020noise} also argues in more details that QAQC exhibits strong noise resilience as observed by us.

\begin{figure}[]
    \captionsetup{justification=raggedright}
    \centering
    \includegraphics[width =0.8 \linewidth]{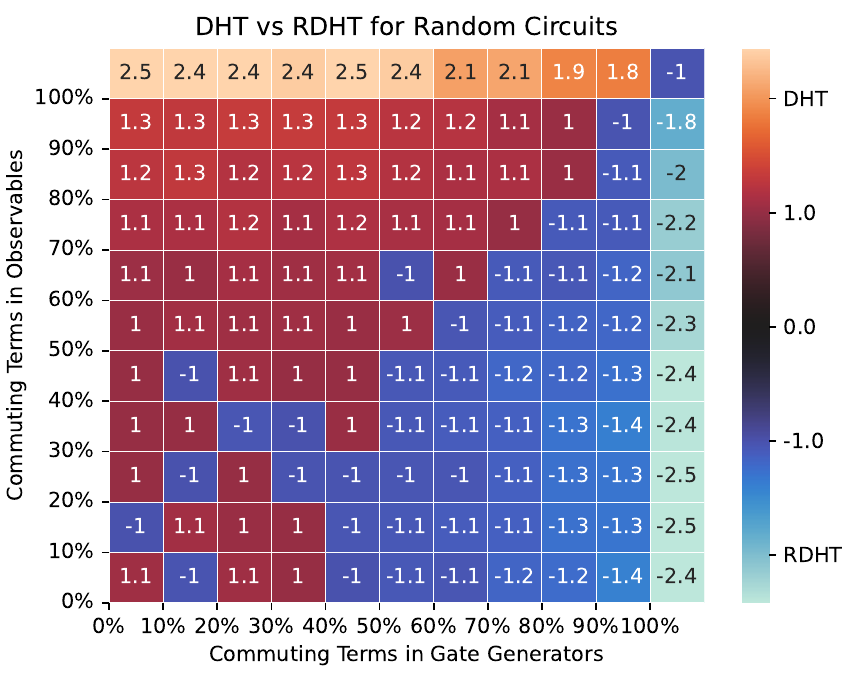}
    \caption{Comparative analysis of DHT and RDHT in terms of the gradient circuits executions for 4-qubit VQAs. On the x-axis, the plot delineates the fraction of commutative terms within the gate generator \(H_j\), while the y-axis quantifies the proportion of mutually commutative terms in the measurement observable \(O\). Each cell in the grid encodes the relative time complexity ratio of DHT to RDHT. This analysis reveals that RDHT gains an efficiency edge when \(H_j\) offers superior groupability as compared to \(O\). In contrast, when grouping of \(O\) is more beneficial, DHT demonstrates superior performance.}
    \label{fig:dht_vs_rdht}
\end{figure}

\remark{As discussed in this section, reversed methods can reduce distinct circuit execution counts when the generator \(H_j\) benefits more from measurement optimization (as described in Sec.~\ref{sec:measurement_grouping}) than the observable \(O\). To further illustrate this, we test the performance of DHT and RDHT on synthetic quantum circuits, varying the potential advantage of measurement optimization techniques for both the generator and observable. The results, shown in Fig.~\ref{fig:dht_vs_rdht}, highlight the significant impact of selecting appropriate gradient methods for different PQCs.}

\subsection{Quantum Neural Network}
\label{sec:power_QAD}

In Section.~\ref{sec:QAD}, we have explained that the power of QAD comes from different treatments for each individual parameter in a PQC.

\begin{figure}[]
    \captionsetup{justification=raggedright}
    \centering
    \includegraphics[width=\linewidth]{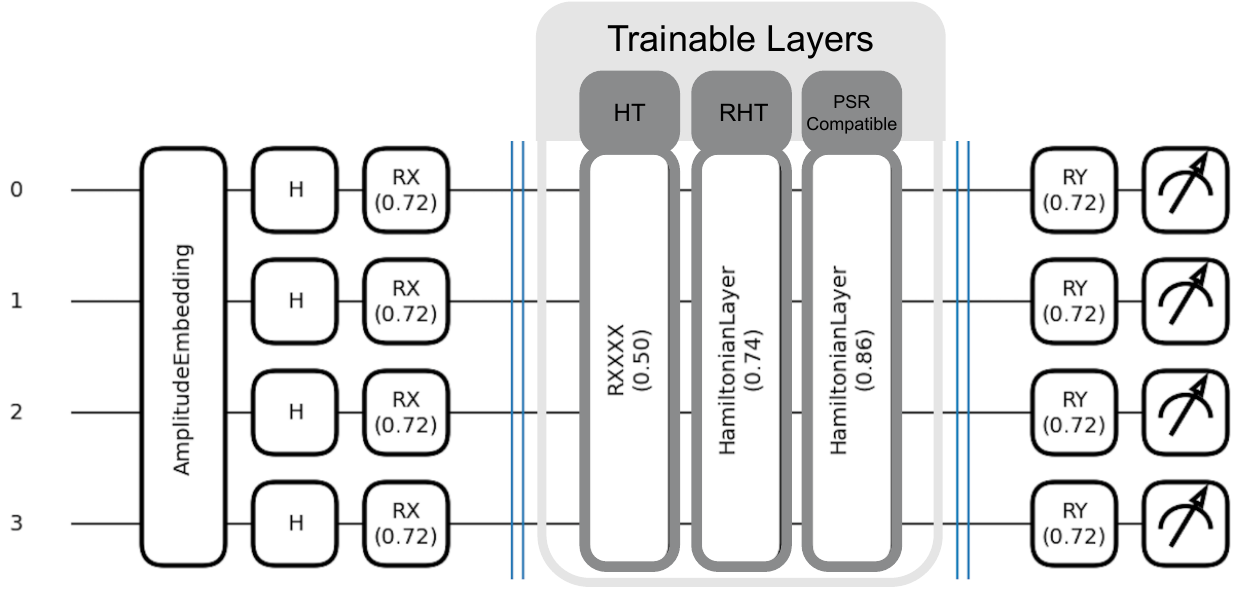}
    \caption{Our handcrafted ansatz has three trainable gates: with generators: $H_1 = XXXX$; $H_2 = \sum_{i}\alpha_i P_i$, where $P_i \in \{I,Z\}^{\otimes 4}$; and $H_3 = \sum_{i} P_i$, where $P_i \in \{I, X\}^{\otimes 4}$. $H_2$ and $H_3$ have 16 terms each, with $U_3$ being PSR-compatible without any decomposition as it only has two distinct eigenvalues. The observable used was $O = Z_0Z_1 + Z_1Z_2 + Z_2Z_3 + Z_3Z_0$, exhibiting full commutativity. While the first gate prefers HT, the second gate prefers RHT, and the third gate prefers PSR}
    \label{fig:qnn_circuit}
\end{figure}

\textbf{Benchmarks:} In this experiment, we handcraft a quantum ansatz which exhibits different structure for different parameters, as shown in Fig.~\ref{fig:qnn_circuit}. We then utilize this synthetic ansatz to train a variational quantum classifier in the Quantum Neural Network (QNN) setting \cite{QNN_power}. The task is to classify a dataset into two distinct classes within the renowned Iris dataset \cite{iris_dataset}. Previously in Appendix.~\ref{sec:appendix_direct}, we evaluated the \textit{circuit fidelity} for each gradient method, while in Section.~\ref{sec:power_reversed}, we assessed the \textit{distinct circuit execution count}. QAD is designed to work with various cost metrics. In this section, we adopt both circuit fidelity and circuit count as metrics and compare QAD with other methods, as done separately in the aforementioned sections.

\begin{figure}[h!]
    \captionsetup{justification=raggedright}
    \centering
    \includegraphics[width=\linewidth]{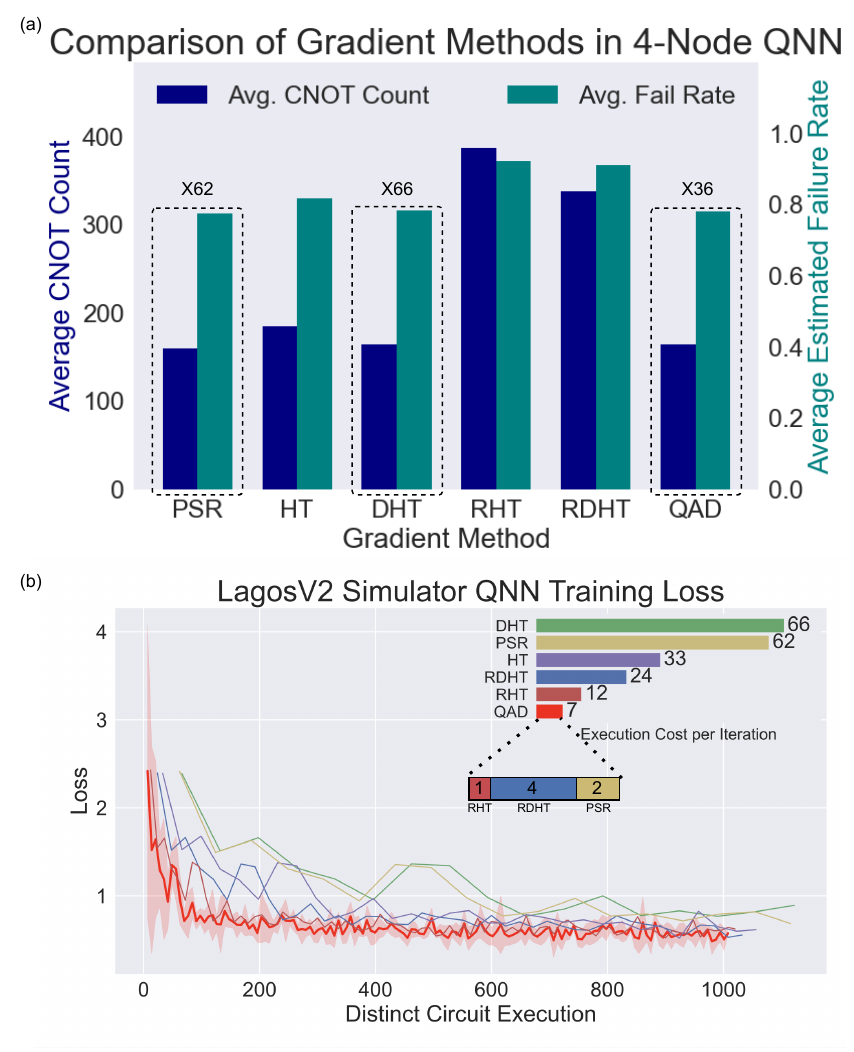}
    \caption{
        \textbf{(a)} The average CNOT counts and gradient circuit EFR when QAD takes the EFR as the cost metric, the numbers above the dashed rectangles are the circuit execution cost per iteration. \textbf{(b)} Now QAD takes the circuit execution count as the cost metric. And the plot is the training loss curve for the QNN classification task with the handcrafted ansatz. The mini bar plot in the top right corner is the execution counts per iteration for each method. 
    }
    \label{fig:qnn}
\end{figure}

\textbf{Baselines:} Note that this handcrafted ansatz is also PSR-compatible after decomposition, so in this experiment we have all the mentioned methods as baselines: PSR, HT, DHT, RHT, RDHT and QAD. More specifically, as illustrated in Fig.~\ref{fig:qnn_circuit}, the QNN circuit has three trainable parameters and the best gradient methods for them are HT, RHT and PSR respectively if circuit count is chosen to be the metric. So QAD will choose 33\% HT, 33\% RHT and 33\% PSR for this task.

\textbf{Results:} In Fig.~\ref{fig:qnn} (a), we see that QAD achieves the best circuit fidelity when the user set the cost metric to be it. And interestingly, three dashed rectangles represent circuit fidelity for the PSR, DHT, and QAD methods, all of which exhibit similar levels of performance. However, even though QAD does not intend to optimize over circuit execution counts, it still achieves at least a $2\times$ speed up compared to the other methods. In Fig.~\ref{fig:qnn} (b), we present the convergence rate for all gradient methods when the metric of QAD is chosen to be the circuit count. For this QNN task, even though PSR can be made compatible after decomposition, it suffers from the huge overhead that \textit{62} distinct circuits need to be executed to obtain the gradients. QAD employs the best treatment for each individual parameter and achieves about $9\times$ speed up than the naive PSR.

%% file: 5_discussion.tex
\section{Related Work}
\label{sec:related_work}
Recent advancements in measurement grouping strategies, as delineated in \cite{grouping_1, grouping_efficient, grouping_graph_theory, grouping_unitary_partioning, measurement_grouping_yongshan}, have shown significant potential in substantially reducing the number of unique circuit evaluations required. These strategies can be effectively applied to the various gradient methods discussed in this work, enhancing the efficiency of our proposed algorithms and further reducing the necessary circuit evaluations for gradient computations.

Beyond grouping commutative terms during quantum circuit measurements, the technique of classical shadows, as proposed in \cite{huangClassicalShadows}, offers a promising avenue for measurement reduction. This approach can potentially scale down the measurement complexity from \(O(M)\) to \(O(\log M)\) for certain classes of observables in the shadow tomography setting \cite{aaronson2018shadow}. Integrating such methodologies into gradient estimation is non-trivial because the distinct gradient circuits for different parameters are not straightforwardly compatible with the shadow tomography setting. A recent paper \cite{abbas2024quantum} established a reduction from gradient estimation to shadow tomography and demonstrated a gradient estimation algorithm with classical-shadow-like scaling in terms of quantum cost, although it overlooks the prohibitive classical cost present in all known shadow tomography schemes. With the reversed methods, where we exchange the roles of unitary gates and measurements, the gradient circuits for different parameters share a common structure but differ in the measurements. This presents an opportunity for future work to establish similar reductions and employ shadow tomography solutions, even though one must endure the substantial classical cost. Moreover, these theories pave the way for novel ansatz designs that facilitate such reductions or are more amenable to classical cost considerations.

Furthermore, our reversed methods themselves inspire new ansatz designs, particularly in scenarios like quantum chemistry, where measurement protocols are inherently task-specific and less flexible. Our reversed methods and QAD provide an opportunity to apply these theories in designing efficient variational ansatz, where the generators, when repurposed as measurements, can leverage various measurement optimization techniques for expedited optimization convergence.

Additionally, the symmetry observed in the structures of (D)HT and R(D)HT prompts an inquiry into the existence of analogous reversed circuit constructs for other prevalent gradient methods.

For the $k$-fold Hadamard Test algorithm, the ability to evaluate higher-order derivatives using the generalized Hadamard test opens up possibilities for extending many existing works, such as efficient gradient methods for PQCs that leverage Lie algebraic symmetries \cite{heidari2024efficient}. We emphasize that the algorithm does not merely evaluate higher-order derivatives but, more broadly, evaluates the expectation value of nested commutators. Moreover, Algorithm \ref{alg:higher_ht} can be easily adapted to evaluate nested combinations of commutators and anti-commutators. We believe this capability could be related to the nested commutators appearing in Trotterization, but we leave the exploration of this potential relationship and its usefulness for future work.

Additionally, note that the costs in Table \ref{tab:k_methods} are for a fixed set of indices. To understand the scaling of these costs, consider the different orders of derivatives: the full information for the first-order gradient is a vector, for the second-order derivatives, it is the Hessian matrix, and for higher-order derivatives, the full information is a tensor of size \(n^k\), where \(n\) is the number of parameters. Therefore, to obtain the full information tensor, the cost needs to be multiplied by \(n^k\).  When \(n \gg k\), the term \(n^k\) will dominate the \(2^k\) terms in Table \ref{tab:k_methods}. \cite{mari2021estimating} discovered that for simple parameterized gates compatible with PSR, the number of distinct circuits to execute does not scale with \(k\); a constant number of circuit executions contains the full information of the higher-order derivatives in such PQCs. This phenomenon is not general because the generators of gates compatible with PSR have only two distinct eigenvalues, restricting the degree of variation of that parameter so that its higher-order derivatives do not contain additional information. Generally, the scaling by \(n^k\) is unavoidable, but we anticipate finding special cases of PQCs where this cost can be mitigated.

%% file: 6_conclusion.tex
\section{Conclusion}
\label{sec:conclusion}

In this study, we addressed the crucial role of gradient calculations for parameterized quantum circuits by introducing efficient quantum gradient methods. Our contributions include the development of a general technique, the Flexible Hadamard Test, which induces two first-order gradient methods, Reversed Hadamard Test and Reversed Direct Hadamard Test. Additionally, we introduced the \(k\)-fold Hadamard Test for evaluating \(k^{th}\)-order partial derivatives using only one quantum circuit, and the aggregated Quantum Automatic Differentiation method. Through numerical evaluations, we validated the efficacy of these methods and explored their comparative performance in diverse contexts.

In tasks such as Quantum-Assisted Quantum Compiling with complex Hamiltonians, our reversed methods demonstrated a reduction in circuit execution counts by a factor of \(O(N)\), where \(N\) represents the number of qubits in the target quantum computer. In Quantum Neural Network classification tasks, QAD adaptively selects the best gradient method for each parameter in the PQC. For this specific QNN ansatz, QAD achieves a \(9\times\) faster convergence than the naive parameter shift rule. This adaptability mirrors the ease brought by classical automatic differentiation systems in computing gradients for optimization and machine learning, paving the way for QAD to facilitate accurate and efficient quantum gradient computation without manual intervention.

Overall, our work advances the field of quantum computing by providing novel, efficient algorithms for gradient and higher-order derivative computation, and an adaptive approach that enhances the performance and efficiency of variational quantum algorithms. These contributions offer valuable insights and practical solutions for various application domains, significantly improving the execution efficiency and convergence rates of quantum algorithms in the NISQ era. By addressing the limitations of existing methods and introducing flexible and adaptive techniques, we pave the way for more robust and scalable quantum computing applications.

%% file: 8_acknowledgements.tex
\begin{acknowledgements}
We would like to thank Shifan Xu and Zhiding Liang for meaningful discussions. This project was supported by the National Science Foundation (under award CCF-2338063) and was developed with funding from the Defense Advanced Research Projects Agency (DARPA). The views, opinions, and/or findings expressed are those of the authors and should not be interpreted as representing the official views or policies of the Department of Defense or the U.S. Government. External interest disclosure: YD is a scientific advisor to, and receives consulting fees from Quantum Circuits, Inc.
\end{acknowledgements}

%% file: 7_appendix.tex
\section{Appendix}
\label{sec:appendix}

\subsection{Direct Hadamard Test}
\label{sec:appendix_DHT}

The Direct Hadamard Test (DHT) proposed in \cite{indirect_to_direct} replaces the indirect measurement in HT with direct measurements. DHT removes the ancillary qubit but doubles the circuits to be executed, as illustrated in Fig.~\ref{fig:grad_methods}(c). DHT estimates the gradient in Eq.~\eqref{eq:simple_grad_def} by:
\begin{align}
\frac{\partial f(\boldsymbol{\theta})}{\partial \theta_j} &= -\frac{1}{2}
    \sum\limits_{k=1}^{N_j}
    {\beta_k^{(j)}
    \left(
        \langle O \rangle_{+,k} - \langle O \rangle_{-,k}
    \right)},
\end{align}
where $\langle O \rangle_{\pm,k}$ is the expected value of the measurement of the plus/minus circuit in Fig.~\ref{fig:grad_methods}(c). In comparison, HT contains controlled multi-qubit gates, which requires the ancillary qubit to interact with many qubits. In NISQ hardwares, such native connectivity is hardly satisfied, so additional SWAP gates will be inserted to implement such gates. This observation implies that DHT, although doubles the circuit execution counts, has better fidelity when executed. We will demonstrate this numerically in Section.~\ref{sec:appendix_direct}

\subsection{Flexible Hadamard Test Decomposition Example}
\label{sec:appendix_flex_HT_example}

In this section, we demonstrate how to apply decomposition before the flexible Hadamard Test (HT) with an example. We set up the problem to be estimating \(\Im{ \langle \prod_{i=1}^{3} \hat{H}_i \rangle_{\ket{\boldsymbol{\theta}}}}\) with a single-qubit system where:

\begin{align*}
    \Bigg\langle \prod_{i=1}^{3} \hat{H}_i \Bigg\rangle_{\ket{\boldsymbol{\theta}}} &= \bra{\boldsymbol{\theta}} \hat{H}_1 \hat{H}_2 \hat{H}_3 \ket{\boldsymbol{\theta}} \\
    \hat{H}_1 = \hat{H}_2 &= \hat{H}_3 = X + Z.
\end{align*}

Note that although in this example one can easily compute \(\hat{H}_1 \hat{H}_2 \hat{H}_3\), in practice, these Hermitian operators are typically Paulis conjugated by unitaries, making it infeasible to compute their product in advance of preparing quantum circuits.

Recall that in Algorithm \ref{alg:pick_ht}, one needs to implement \(\hat{H}_i\) as part of a controlled gate. However, when each individual \(\hat{H}_i\) is not easily implemented as unitary gates, we can decompose such terms into linear combinations of Pauli words. Here, we set their Pauli decomposition to be \(X + Z\), which is Hermitian. To demonstrate Algorithm \ref{alg:pick_ht}, without loss of generality, we choose to measure \(\hat{H}_2\). Thus,

\begin{align}
    \bra{\boldsymbol{\theta}} \hat{H}_1 \hat{H}_2 \hat{H}_3 \ket{\boldsymbol{\theta}} &= \bra{\boldsymbol{\theta}} (X+Z) \hat{H}_2 (X+Z) \ket{\boldsymbol{\theta}} \nonumber \\
    &= \bra{\boldsymbol{\theta}} X \hat{H}_2 X \ket{\boldsymbol{\theta}} + \bra{\boldsymbol{\theta}} X \hat{H}_2 Z \ket{\boldsymbol{\theta}} \nonumber \\
    &\quad + \bra{\boldsymbol{\theta}} Z \hat{H}_2 X \ket{\boldsymbol{\theta}} + \bra{\boldsymbol{\theta}} Z \hat{H}_2 Z \ket{\boldsymbol{\theta}}.
    \label{eq:appendix_flex_HT_decomposition}
\end{align}

Now, each of the four terms in Eq.~\eqref{eq:appendix_flex_HT_decomposition} can be estimated by Algorithm \ref{alg:pick_ht} since it is straightforward to implement controlled Pauli gates. This decomposition approach is applicable when using HT, DHT, RHT, or RDHT for parameterized circuits with complex gate generators \(\{H_j\}_j\).

\begin{remark}
To implement \( \text{controlled-}UPU^\dagger \), where \(P\) is a Pauli operator conjugated by unitaries as in Eq.~\eqref{eq:def_herm_op}, one can achieve this by first applying an unconditional \(U\), followed by \(\text{controlled-}P\), and then an unconditional \(U^\dagger\).
\end{remark}

\subsection{Detailed Flexible Hadamard Test Explanation}
\label{sec:appendix_flex_HT}

\begin{figure}[h]
    \captionsetup{justification=raggedright}
    \centering
    \includegraphics[width=\linewidth]{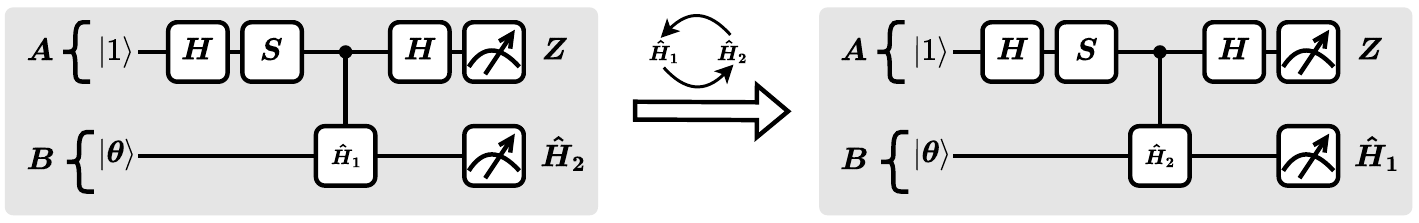}
    \caption{Swap \(\boldsymbol{\hat{H}_{1}}\) and \(\boldsymbol{\hat{H}_{2}}\) in Hadamard Test}
    \label{fig:HT_swap}
\end{figure}

To evaluate \(\Im{\bra{\boldsymbol{\theta}} \hat{H}_1 \hat{H}_2 \ket{\boldsymbol{\theta}}}\), one can use a Hadamard test circuit as shown in the left circuit of Fig.~\ref{fig:HT_swap}. In this setup, \(\hat{H}_1\) is implemented as a controlled gate, while \(\hat{H}_2\) is measured at the end. Due to the symmetry of \(\hat{H}_1\) and \(\hat{H}_2\) in \(\Im{\bra{\boldsymbol{\theta}} \hat{H}_1 \hat{H}_2 \ket{\boldsymbol{\theta}}}\), swapping the roles of \(\hat{H}_1\) and \(\hat{H}_2\) in the Hadamard test circuit will compute the desired value up to a sign change. This swap results in the right circuit of Fig.~\ref{fig:HT_swap}, where \(\hat{H}_1\) is measured instead of \( \hat{H}_2\), allowing for the computation of the target value with the roles of \(\hat{H}_1\) and \(\hat{H}_2\) reversed.

With the measurement grouping techniques discussed in Sec.~\ref{sec:measurement_grouping}, the computational complexity of these two circuits varies depending on the specific instantiation of \(\hat{H}_1\) and \(\hat{H}_2\).

\begin{figure}[h]
    \captionsetup{justification=raggedright}
    \centering
    \includegraphics[width=\linewidth]{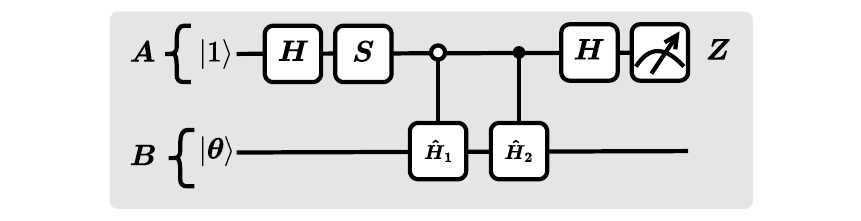}
    \caption{Empty-controlled Gates in Hadamard Test for estimating \(\Im{\bra{\boldsymbol{\theta}} \hat{H}_1 \hat{H}_2 \ket{\boldsymbol{\theta}}}\)}
    \label{fig:HT_empty}
\end{figure}

A natural question arises: does this approach generalize to cases where the product involves more than two terms? For example, if one needs to evaluate \(\Im{\bra{\boldsymbol{\theta}} \hat{H}_1 \hat{H}_2 \hat{H}_3 \ket{\boldsymbol{\theta}}}\) and wishes to measure \(\hat{H}_2\), the circuits in Fig.~\ref{fig:HT_swap} are not applicable. Instead, circuits designed to measure \(\hat{H}_2\) would yield either \(\Im{\bra{\boldsymbol{\theta}} \hat{H}_1 \hat{H}_3 \hat{H}_2 \ket{\boldsymbol{\theta}}}\) or \(\Im{\bra{\boldsymbol{\theta}} \hat{H}_2 \hat{H}_1 \hat{H}_3 \ket{\boldsymbol{\theta}}}\). However, it is important to observe that empty-controlled gates mathematically contribute to the opposite side of the product compared to solid-controlled gates, as illustrated in Fig.~\ref{fig:HT_empty}.

The circuit in Fig.~\ref{fig:HT_empty} is also known for implementing a SWAP test with a Hadamard test. By combining the circuits in Fig.~\ref{fig:HT_swap} and Fig.~\ref{fig:HT_empty}, we introduce the flexible Hadamard test circuit shown in Fig.~\ref{fig:flexible_HT}. This new circuit allows for the measurement of any arbitrary \(\hat{H}_i\) to compute \(\Im{\bra{\boldsymbol{\theta}}\hat{H}_1\cdots \hat{H}_m \ket{\boldsymbol{\theta}}}\). The correctness of the flexible Hadamard test algorithm, as described in Algorithm~\ref{alg:pick_ht}, is demonstrated with the following details:

\begin{align*}
    &Initial\_State = \ket{1} \otimes \ket{\boldsymbol{\theta}} \\
    \xrightarrow{H_0S_0}& \frac{1}{\sqrt{2}} \left ( \ket{0} \otimes \ket{\boldsymbol{\theta}} - i\ket{1} \otimes \ket{\boldsymbol{\theta}} \right) \\
    \xrightarrow{C_0-\hat{H}_1\cdots \hat{H}_{i-1}}& \frac{1}{\sqrt{2}}  ( \ket{0} \otimes \hat{H}_{i-1}\cdots \hat{H}_1\ket{\boldsymbol{\theta}} -i \ket{1} \otimes \ket{\boldsymbol{\theta}} ) \\
    \xrightarrow{C-\hat{H}_m\cdots \hat{H}_{i+1}}& \frac{1}{\sqrt{2}} \bigg ( \ket{0} \otimes \hat{H}_{i-1}\cdots \hat{H}_1\ket{\boldsymbol{\theta}}\\
    & \quad -i \ket{1} \otimes \hat{H}_{i+1}\cdots \hat{H}_{m}\ket{\boldsymbol{\theta}} \bigg) 
\end{align*}

After measuring against $X_0\otimes \hat{H}_i$,

\begin{align*}
    \langle X_0\otimes \hat{H}_i \rangle =& \frac{i}{2}\biggl ( \bra{\boldsymbol{\theta}}\hat{H}_m\cdots \hat{H}_{i+1} \hat{H}_i \hat{H}_{i-1} \cdots \hat{H}_1 \ket{\boldsymbol{\theta}} \\
    &- \bra{\boldsymbol{\theta}}\hat{H}_1\cdots \hat{H}_{i-1} \hat{H}_i \hat{H}_{i+1} \cdots \hat{H}_m \ket{\boldsymbol{\theta}}\biggl ) \\
    =& \Im{\bra{\boldsymbol{\theta}}\hat{H}_1\cdots \hat{H}_m \ket{\boldsymbol{\theta}}} \\
    =& \Im{\langle \prod_{i=1}^{m} \hat{H}_i \rangle_{\ket{\boldsymbol{\theta}}}}
\end{align*}

This computes the imaginary part of the expected value of an \(m\)-term product as defined in Def.~\ref{def:m-term_product_exp_value}. To compute the real part, one can simply change the initial state of the ancillary qubit.

\subsection{Proof of Reversed Hadamard Test Correctness}
\label{sec:appendix_RHT}

To verify the correctness of RHT, we mathematically derive the measurement outcome of the RHT circuit and verify it is equivalent to Eq.~\eqref{eq:grad_def_com}

\begin{align*}
    &Initial\_State = \ket{1} \otimes \ket{\Psi} \\
    \xrightarrow{(H_0S_0) \otimes U_{1:n}}& \frac{1}{\sqrt{2}} \left ( \ket{0} \otimes U_{1:n}\ket{\Psi} - i\ket{1} \otimes U_{1:n}\ket{\Psi} \right) \\
    \xrightarrow{C-P_l}& \frac{1}{\sqrt{2}} \left ( \ket{0} \otimes U_{1:n}\ket{\Psi} -i \ket{1} \otimes P_lU_{1:n}\ket{\Psi} \right) \\
    \xrightarrow{I_0 \otimes U_{j+1:n}^\dagger}& \frac{1}{2} ( \ket{0} \otimes U_{1:j} \ket{\Psi} -  i\ket{1} \otimes U_{j+1:n}^\dagger P_lU_{1:n}\ket{\Psi} ) 
\end{align*}

After measuring against $X_0\otimes H_j$,

\begin{align*}
    \langle X_0\otimes H_j \rangle =& \frac{i}{2}\biggl ( \bra{\Psi}U_{1:j}^\dagger H_jU_{j+1:n}^\dagger P_lU_{1:n}\ket{\Psi} \\
    &- \bra{\Psi}U_{1:n}^\dagger P_lU_{j+1:n}H_jU_{1:j}\ket{\Psi}\biggl ) \\
    =& -\Im{\bra{\Psi}U_{1:j}^\dagger H_jU_{j+1:n}^\dagger P_lU_{1:n}\ket{\Psi}}
\end{align*}

The summation over $l$ is exactly equivalent to Eq.~\eqref{eq:grad_def_com}.

\subsection{Proof of Reversed Direct Hadamard Test Correctness}
\label{sec:appendix_RDHT}

Algorithm~\ref{alg:pick_ht} can be generalized for other hadamard-test-based algorithms. One such example is the DHT in Sec.~\ref{sec:DHT} and Appendix.~\ref{sec:appendix_DHT}. Similar to RHT, we propose the Reversed Direct Hadamard Test (RDHT) to reverse the roles of $H_j$ and $O$, as illustrated in Fig.~\ref{fig:grad_methods}(e). And the gradient is given by the following equation:

\begin{equation}
    \frac{\partial f(\boldsymbol{\theta})}{\partial \theta_j} = \frac{1}{2}
    \sum\limits_{l=1}^{M}
    {\alpha_l
    \left(
        \langle 
            H_j \rangle_+^{(l)} - \langle H_j \rangle_-^{(l)}
    \right)}
\end{equation}
where $\langle H_j \rangle_+^{(l)}$ is the plus circuit in Fig.~\ref{fig:grad_methods}(e) and $\langle H_j \rangle_-^{(l)}$ is the minus circuit. The proof of correctness could be found in Appendix.~\ref{sec:appendix_RDHT}

RDHT merges the characteristics of DHT and RHT, as elaborated in the preceding sections and the numerical experiment segment of this paper. The four methods HT, DHT, RHT, and RDHT demonstrate varying strengths and weaknesses across various variational algorithms and even for different parameters within one VQA, attributable to the distinct nature of their generators $\{H_j\}_j$ and observable $O$.

Similar to Appendix.~\ref{sec:appendix_RHT}, we can verify the correctness of RDHT by evaluating the plus and minus circuits. Let us start with the plus circuit in Fig.~\ref{fig:grad_methods}(e):

\begin{align*}
    &Initial\_State = \ket{\Psi} \\
    \xrightarrow{U_{1:n}}& U_{1:n}\ket{\Psi} \\
    \xrightarrow{e^{+i\frac{\pi}{4}P_l} = \frac{1}{\sqrt{2}} (I + iP_l)}& \frac{1}{\sqrt{2}} \biggl ( U_{1:n} \ket{\Psi} + iP_lU_{1:n} \ket{\Psi} \biggl ) \\
    \xrightarrow{U_{j+1, n}^\dagger}& \frac{1}{\sqrt{2}} \biggl ( U_{1:j} \ket{\Psi} + iU_{j+1: n}^\dagger P_lU_{1:n} \ket{\Psi} \biggl )
\end{align*}

After measuring against $H_j$,

\begin{align*}
    \langle H_j \rangle_+ =& \frac{1}{2} \biggl ( \bra{\Psi}U_{1:j}^\dagger H_jU_{1:j}\ket{\Psi} \\
    &- \bra{\Psi}U_{1:n}^\dagger P_lU_{j+1:n}H_jU_{j+1:n}^\dagger P_l U_{1:n}\ket{\Psi} \\
    &+i \bra{\Psi}U_{1:j}^\dagger H_jU_{j+1:n}^\dagger P_lU_{1:n}\ket{\Psi} \\
    &+i \bra{\Psi}U_P{1:n}^\dagger P_lU_{j+1:n}H_jU_{1:j}\ket{\Psi} \biggl )
\end{align*}

Likewise,
\begin{align*}
    \langle H_j \rangle_- =& \frac{1}{2} \biggl ( \bra{\Psi}U_{1:j}^\dagger H_jU_{1:j}\ket{\Psi} \\
    &- \bra{\Psi}U_{1:n}^\dagger P_lU_{j+1:n}H_jU_{j+1:n}^\dagger P_l U_{1:n}\ket{\Psi} \\
    &-i \bra{\Psi}U_{1:j}^\dagger H_jU_{j+1:n}^\dagger P_lU_{1:n}\ket{\Psi} \\
    &-i \bra{\Psi}U_P{1:n}^\dagger P_lU_{j+1:n}H_jU_{1:j}\ket{\Psi} \biggl )
\end{align*}
And the difference between these two
\begin{align*}
    \frac{1}{2} (\langle H_j \rangle_+ - \langle H_j \rangle_-) = -\Im{\bra{\Psi}U_{1:j}^\dagger H_jU_{j+1:n}^\dagger P_lU_{1:n}\ket{\Psi}}
\end{align*}
The summation over $l$ is exactly equivalent to Eq.~\eqref{eq:grad_def_com}.

\subsection{Higher-order Derivative of PQC}
\label{sec:appendix_k_th_order_value}

For objective function $f$ defined in Eq.~\eqref{eq:f(theta)}:
\begin{align*}
     f(\boldsymbol{\theta}) = \bra{\Psi}U_{1:n}^\dagger O U_{1:n}\ket{\Psi} = \bra{\boldsymbol{\theta}}O\ket{\boldsymbol{\theta}}
\end{align*}
We show by induction that for \( 1 \leq j_1 \leq j_2 \leq \cdots \leq j_k \leq n \), the k-th order partial derivative is:
\begin{align*}
    &\frac{\partial^k f}{\partial \theta_{j_1} \partial \theta_{j_2} \cdots \partial \theta_{j_k}} (\boldsymbol{\theta})  \nonumber \\
    =\quad & \left(\frac{i}{2}\right)^k \bra{\boldsymbol{\theta}} [\tilde{H}_{j_1}, [\tilde{H}_{j_2}, [\cdots, [\tilde{H}_{j_k}, O]]]] \ket{\boldsymbol{\theta}}
\end{align*}
\textbf{Base Case:} For \( k = 1 \), the first-order derivative or gradient is used by many papers \cite{harrow2021low} and is easy to verify that:
\begin{align*}
    \frac{\partial f}{\partial \theta_{j_1}} = \frac{i}{2}\bra{\boldsymbol{\theta}} [\tilde{H}_{j_1}, O] \ket{\boldsymbol{\theta}}
\end{align*}
\noindent\textbf{Inductive Step:} Assume the claim is true for some \( k \geq 1 \). Specifically:
\begin{align*}
    &\frac{\partial^k f}{\partial \theta_{j_1} \partial \theta_{j_2} \cdots \partial \theta_{j_k}} (\boldsymbol{\theta})  \nonumber \\
    =\quad & \left(\frac{i}{2}\right)^k \bra{\boldsymbol{\theta}} [\tilde{H}_{j_1}, [\tilde{H}_{j_2}, [\cdots, [\tilde{H}_{j_k}, O]]]] \ket{\boldsymbol{\theta}}
\end{align*}
We need to show that for $k+1$ indices \( 1 \leq j_1 \leq j_2 \leq \cdots \leq j_k \leq j_{k+1} \leq n \):
\begin{align*}
    &\frac{\partial^k f}{\partial \theta_{j_1}  \partial \theta_{j_2} \cdots \partial \theta_{j_k} \partial \theta_{j_{k+1}}} (\boldsymbol{\theta})  \nonumber \\
    =\quad & \left(\frac{i}{2}\right)^{k+1} \bra{\boldsymbol{\theta}} [\tilde{H}_{j_1},[\tilde{H}_{j_2}, [\cdots, [\tilde{H}_{j_k}, [\tilde{H}_{j_{k+1}}, O]]]] \ket{\boldsymbol{\theta}}
\end{align*}

Note that for the $k$ indices \( 1 \leq j_2 \leq j_3 \leq \cdots \leq j_k \leq j_{k+1} \leq n \), by inductive hypothesis we have:
\begin{align*}
    g(\boldsymbol{\theta}) =\quad & \frac{\partial^k f}{\partial \theta_{j_2} \partial \theta_{j_3} \cdots \partial \theta_{j_{k+1}}} (\boldsymbol{\theta}) \nonumber \\
    =\quad & \left(\frac{i}{2}\right)^k \bra{\boldsymbol{\theta}} [\tilde{H}_{j_2}, [\tilde{H}_{j_3}, [\cdots, [\tilde{H}_{j_{k+1}}, O]]]] \ket{\boldsymbol{\theta}} \\
    \coloneqq \quad & \left(\frac{i}{2}\right)^k \bra{\boldsymbol{\theta}} \mathbf{M} \ket{\boldsymbol{\theta}}
\end{align*}
where we denote $[\tilde{H}_{j_2}, [\tilde{H}_{j_3}, [\cdots, [\tilde{H}_{j_{k+1}}, O]]]]$ as $\mathbf{M}$ and
\begin{align*}
    \frac{\partial^k f}{\partial \theta_{j_1}  \partial \theta_{j_2} \cdots \partial \theta_{j_k} \partial \theta_{j_{k+1}}} (\boldsymbol{\theta}) = \frac{\partial g}{\partial \theta_{j_1}} (\boldsymbol{\theta})
\end{align*}
Notice that $\mathbf{M}$ does not depend on $\theta_{j_1}$ since \( 1 \leq j_1 \leq j_2 \leq j_3 \leq \cdots \leq j_k \leq j_{k+1} \leq n \), so we have:
\begin{align*}
    &\quad \frac{\partial g}{\partial \theta_{j_1}} (\boldsymbol{\theta}) \\
    &= \left(\frac{i}{2}\right)^k \Bigg( \bra{\Psi} U_1^\dagger \cdots U_{j_1}^\dagger (+iH_{j_1}/2) U_{j_1+1}^\dagger \cdots U_{n}^\dagger \mathbf{M} U_n \cdots U_1 \ket{\Psi} \\
    &\quad  + \bra{\Psi} U_1^\dagger \cdots U_n^\dagger \mathbf{M} U_n \cdots U_{j_1+1} (-iH_{j_1}/2) U_{j_1} \cdots U_1 \ket{\Psi} \Bigg) \\
    &= \left(\frac{i}{2}\right)^k \Bigg( \bra{\boldsymbol{\theta}} \frac{i}{2} \tilde{H}_{j_1} \mathbf{M} \ket{\boldsymbol{\theta}} - \bra{\boldsymbol{\theta}} \mathbf{M} \frac{i}{2} \tilde{H}_{j_1} \ket{\boldsymbol{\theta}} \Bigg)\\
    &= \left(\frac{i}{2}\right)^{k+1} \Bigg( \bra{\boldsymbol{\theta}} [\tilde{H}_{j_1}, \mathbf{M}] \ket{\boldsymbol{\theta}} \Bigg)
\end{align*}
so if we substitute the definition for $g$ and $\mathbf{M}$ we get:
\begin{align*}
    &\frac{\partial^k f}{\partial \theta_{j_1}  \partial \theta_{j_2} \cdots \partial \theta_{j_k} \partial \theta_{j_{k+1}}} (\boldsymbol{\theta})  \nonumber \\
    =\quad & \left(\frac{i}{2}\right)^{k+1} \bra{\boldsymbol{\theta}} [\tilde{H}_{j_1},[\tilde{H}_{j_2}, [\cdots, [\tilde{H}_{j_k}, [\tilde{H}_{j_{k+1}}, O]]]] \ket{\boldsymbol{\theta}}
\end{align*}

\subsection{DHT for $k^{th}$-order Partial Derivative Estimation}
\label{sec:appendix_DHT_for_k}

Besides the naive usage of DHT to replace HT as originally stated in \cite{indirect_to_direct}, we propose a new algorithm for using DHT to directly evaluate the \(k^{th}\)-order derivative. As shown in Table \ref{tab:k_methods}, HT requires \(2^{k-1}\) circuits to evaluate the gradient, with each circuit involving \(k\) controlled gates. To use direct measurements rather than indirect measurements, \(2^k\) direct circuits are needed for each indirect circuit with \(k\) controlled gates. This results in \(2^{k-1} \cdot 2^k\) direct circuits for evaluating the \(k^{th}\)-order derivative. However, we found that DHT can evaluate the nested commutators in Eq.~\eqref{eq:k-th_derivative} much more efficiently. The algorithm is presented below:

\begin{algorithm}
\caption{DHT for $k^{th}$-order Derivative Estimation}\label{alg:higher_dht}
\begin{algorithmic}[1]
    \State $grad \gets 0$
    \For{each $(s_1, s_2, \ldots, s_k)$ in $\{-1, +1\}^{\otimes k}$}
        \State Implement circuit in Fig.~\ref{fig:dht_for_k} with $(s_1, s_2, \ldots, s_k)$
        \State Obtain measurement outcome $m$
        \State $grad \pluseq \text{sign}(s_1 s_2 \cdots s_k) \times m$
    \EndFor
    \State \Return $grad$ as the estimate for the $k^{th}$-order partial derivative
\end{algorithmic}
\end{algorithm}

\begin{figure}[h!]
    \captionsetup{justification=raggedright}
    \centering
    \includegraphics[width=\linewidth]{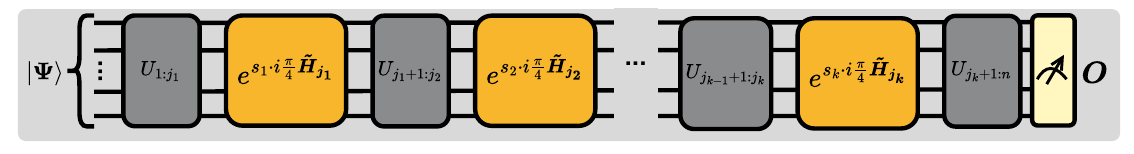}
    \caption{DHT for $k^{th}$-order Derivative Estimation}
    \label{fig:dht_for_k}
\end{figure}

Thus, one can use \(2^k\) DHT circuits to evaluate the \(k^{th}\)-order derivative, as shown in Table \ref{tab:k_methods}.

\subsection{Proof of Correctness for Higher-order Derivative Estimation Algorithm}
\label{sec:appendix_k_th_order_alg}
Effectively we want to show that Algorithm.\ref{alg:higher_ht} as illustrated in Fig.\ref{fig:k-th_order_gradient_algo} outputs an unbiased estimator for the $k$-th order derivative as in Eq.~\eqref{eq:k-th_derivative} for all \( k \geq 1 \). Here we prove it by induction. Denote the quantum state before the last layer of Hadamard gates (before the Pauli-X measurements) in Algorithm.~\ref{alg:higher_ht} as $\ket{\Phi}_{k}$, where $k$ is the number of ancillary qubits and also number of the other Hermitian operators. Specifically, we will prove for any $k+1$ Hermitian operators $\tilde{H}_1, \tilde{H}_2, \cdots, \tilde{H}_{k}, O$:
\begin{align}
    \big\langle X^{\otimes k} \otimes O \big\rangle_{\ket{\Phi}_{k}} =  \left(\frac{i}{2}\right)^k \bra{\boldsymbol{\theta}} [\tilde{H}_{1}, [\tilde{H}_{2}, [\cdots, [\tilde{H}_{k}, O]]]] \ket{\boldsymbol{\theta}}
    \label{eq:appendix_k_order_algo_induction_step}
\end{align}
\textbf{Base Case:} For \( k = 1 \), the algorithm is the standard HT gradient estimation, so it holds for \( k = 1 \).

\noindent\textbf{Inductive Step:} Assume the claim is true for some \( k \geq 1 \). Specifically, for any $k+1$ Hermitian operators $\tilde{H}_1, \tilde{H}_2, \cdots, \tilde{H}_{k}, O$, Eq.~\eqref{eq:appendix_k_order_algo_induction_step} holds.
We need to show for any $k+2$ Hermitian operators $\tilde{H}_1, \tilde{H}_2, \cdots, \tilde{H}_{k+1}, O$:
\begin{align*}
    &\big\langle X^{\otimes k+1} \otimes O \big\rangle_{\ket{\Phi}_{k+1}} = \\
    =\quad & \left(\frac{i}{2}\right)^{k+1} \bra{\boldsymbol{\theta}} [\tilde{H}_{1}, [\tilde{H}_{2}, [\cdots, [\tilde{H}_{k+1}, O]]]] \ket{\boldsymbol{\theta}}
\end{align*}

\begin{figure}[b]
    \captionsetup{justification=raggedright}
    \centering
    \includegraphics[width=\linewidth]{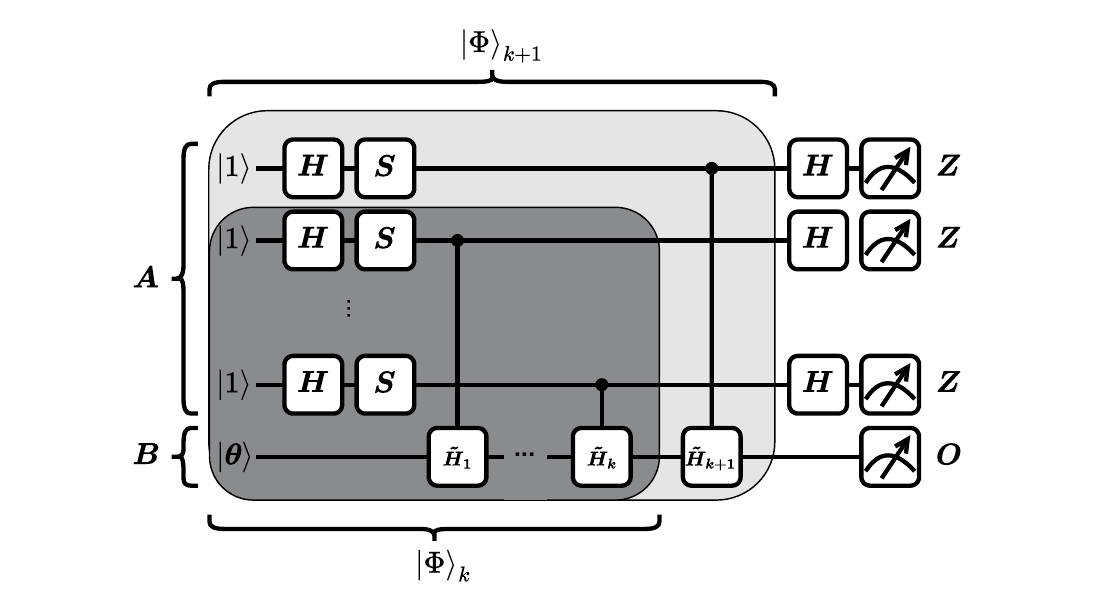}
    \caption{Induction relationship between $(k)$ and $(k+1)$ in higher-order derivative estimation algorithm}
    \label{fig:appendix_k_order_proof_induction}
\end{figure}

Note that $\ket{\Phi}_{k+1}$ has one additional qubit and one controlled gate before measurement, compared to $\ket{\Phi}_{k}$, as illustrated in Fig.~\ref{fig:appendix_k_order_proof_induction}. Intuitively, the induction is based on the fact that $\ket{\Phi}_{k+1}$ is evolved from $\ket{\Phi}_{k}$, and one can observe that:
\begin{align}
    \ket{\Phi}_{k+1} = \bigg (C-\tilde{H}_{k+1}\bigg ) \left( \frac{1}{\sqrt{2}}(\ket{0} - i\ket{1})\otimes \ket{\Phi}_{k} \right)
\end{align}
Thus we can obtain:
\begin{align}
    &\big\langle X^{\otimes k+1} \otimes O \big\rangle_{\ket{\Phi}_{k+1}} \nonumber\\
    =\quad & \bra{\Phi}_{k+1} X^{\otimes k+1}\otimes O \ket{\Phi}_{k+1} \nonumber\\
    =\quad & -\frac{i}{2} \left( \bra{\Phi}_{k} X^{\otimes k}\otimes O \cdot I^{\otimes k}\otimes\tilde{H}_{k+1}\ket{\Phi}_{k} \right. \nonumber\\
    &- \left. \bra{\Phi}_{k} I^{\otimes k}\otimes\tilde{H}_{k+1} \cdot X^{\otimes k}\otimes O \ket{\Phi}_{k}\right) \nonumber\\
    =\quad & -\frac{i}{2} \left( \bra{\Phi}_{k} X^{\otimes k}\otimes [O, \tilde{H}_{k+1}] \ket{\Phi}_{k} \right) \nonumber\\
    =\quad & \frac{1}{2} \left( \bra{\Phi}_{k} X^{\otimes k}\otimes i[\tilde{H}_{k+1}, O] \ket{\Phi}_{k} \right)
    \label{eq:appendix_k+1}
\end{align}
Note that the inductive hypothesis Eq.~\eqref{eq:appendix_k_order_algo_induction_step} is true for any $O$ that is a Hermitian operator and $i[\tilde{H}_{k+1}, O]$ is Hermitian. So if we plug it in Eq.~\eqref{eq:appendix_k_order_algo_induction_step} and substitute it back to Eq.~\eqref{eq:appendix_k+1}, we obtain:
\begin{align}
    &\big\langle X^{\otimes k+1} \otimes O \big\rangle_{\ket{\Phi}_{k+1}} \nonumber \\ 
    =\quad & \frac{1}{2}\left( \left(\frac{i}{2}\right)^k \bra{\boldsymbol{\theta}} [\tilde{H}_{1}, [\tilde{H}_{2}, [\cdots, [\tilde{H}_{k}, i[\tilde{H}_{k+1}, O]]]]] \ket{\boldsymbol{\theta}} \right) \nonumber \\ 
    =\quad & \left(\frac{i}{2}\right)^{k+1} \bra{\boldsymbol{\theta}} [\tilde{H}_{1}, [\tilde{H}_{2}, [\cdots, [\tilde{H}_{k+1}, O]]]] \ket{\boldsymbol{\theta}}
\end{align}

Thus, for any $k+2$ Hermitian operators $\tilde{H}_1, \tilde{H}_2, \cdots, \tilde{H}_{k+1}, O$, $\langle X^{\otimes k+1} \otimes O \rangle_{\ket{\Phi}_{k+1}}$ is exactly what we want, completing the inductive step.
Since both the base case and the inductive step have been proven, by the principle of mathematical induction, Algorithm.~\ref{alg:higher_ht} provides an unbiased estimator for the $k$-th order derivative.

\subsection{Numerical Experiment on QAOA for MaxCut}
\label{sec:appendix_direct}

\begin{figure*}[t]
    \captionsetup{justification=raggedright}
    \centering
    \includegraphics[width=\textwidth]{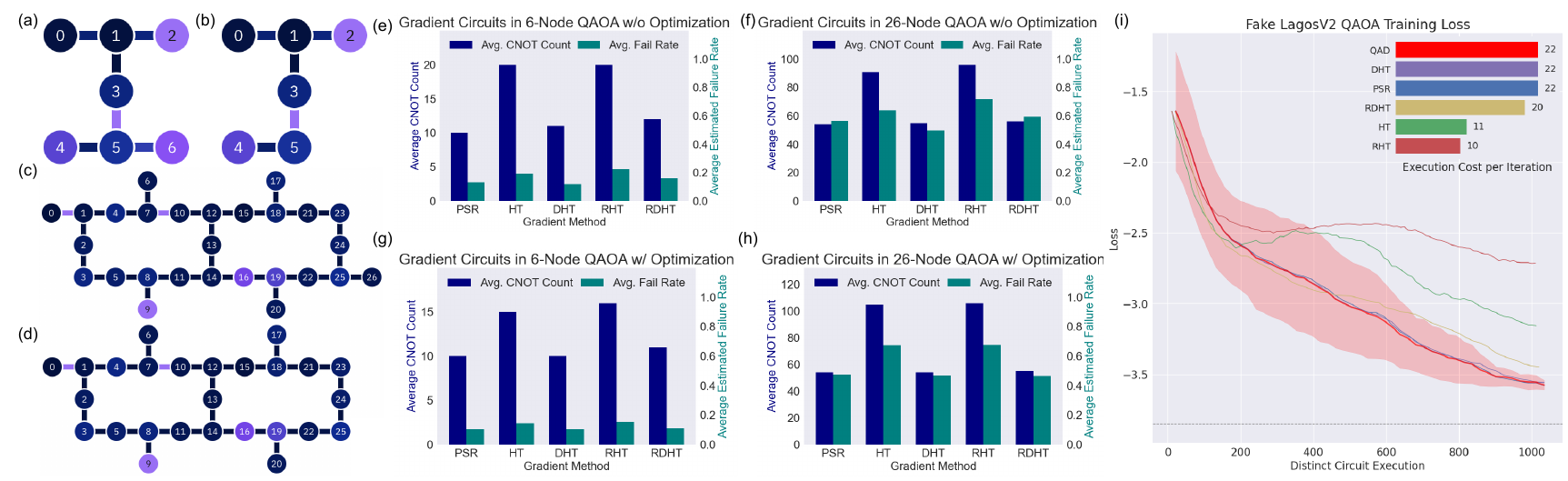}
    \caption{
    \textbf{(a)} the \textit{ibm\_lagos} topology. \textbf{(b)} the Max Cut graph cropped from the hardware topology. \textbf{(c)} the \textit{ibm\_cairo} topology. \textbf{(d)} the Max Cut graph cropped from the hardware topology. \textbf{(e)} The quality analysis for 6 node QAOA without circuit optimization \textbf{(f)} The quality analysis for 26 node QAOA without circuit optimization \textbf{(g)} The quality analysis for 6 node QAOA with Qiskit transpiler \textit{OPT\_LEVEL}=3 \textbf{(h)} The quality analysis for 26 node QAOA with Qiskit transpiler \textit{OPT\_LEVEL}=3 \textbf{(i)} The training convergence for 6 node QAOA without circuit optimization.
    }
    \label{fig:direct_power}
\end{figure*}

\textbf{Benchmarks:} As motivated in Section.~\ref{sec:DHT}, with the cost of doubling the circuit execution count, direct methods (DHT \& RDHT) construct gradient circuits with higher fidelity due to fewer CNOT gates in each circuit. In this experiment, we choose the \textbf{Q}uantum \textbf{A}pproximate \textbf{O}ptimization \textbf{A}lgorithm (QAOA) \cite{QAOA} for MaxCut problem. QAOA is one of the most well-established hybrid quantum-classical variational quantum algorithm designed for solving Quadratic Unconstrained Binary Optimization (QUBO) problems, which are known to be NP-hard. In our setting, the underlying graphs for Max Cut are chosen to be hardware-efficient (i.e. close to the actual connectivity topology of the quantum hardware). We use 7-qubit \textit{ibm\_lagos} (Fig.~\ref{fig:direct_power}(a)) and 27-qubit \textit{ibm\_cairo} (Fig.~\ref{fig:direct_power}(c)) quantum hardwares, and based on the topology of the machines, choose subgraphs (Fig.~\ref{fig:direct_power}(b)(d)) of them as the underlying MaxCut problem. We run QAOA algorithm with different gradient estimation methods, and plot the training loss curves. We also analyze the quality of the gradient circuits within one iteration for all the methods. The quality is measured by average \textit{CNOT Count} and average \textit{Estimated Failure Rate (EFR)}. \textit{EFR} can be efficiently calculated given the circuit and the calibrated error rates of the quantum hardware. Mathematically, $EFR = 1 - \prod_i (1 - p_i)$, where $p_i$ is the error rate for each gate. For each iteration, a gradient method executes a batch of quantum circuits to estimate the gradients. The average \textit{EFR} is calculated across all gradient circuits.

\textbf{Baselines:} We use different gradient methods for this experiment, including PSR, HT, DHT, RHT and RDHT. The direct methods are DHT and RDHT, which should be compared to their non-direct alternative HT and RHT respectively. In QAOA for MaxCut, PSR is equivalent to DHT. In this task, QAD will \textbf{always} choose PSR for different parameters because of its circuits have best quality.

\textbf{Results:} To examine the quality of circuits needed by different gradient methods, we present Fig.~\ref{fig:direct_power} (e)(f)(g)(h). Fig.~\ref{fig:direct_power} (e)(f) show that for the 7-qubit and 27-qubit machine, direct methods exhibit a $2\times$ better average CNOT counts than the non-direct methods, and the average \textit{EFR} for the direct methods are advantageous. Additionally, note that our proposed gradient methods and framework are compatible with arbitrary circuit optimization techniques \cite{haner2020assertion, xu2022quartz, li2022paulihedral}, and we also perform the quality analysis with maximum Qiskit optimization turned on. Fig.~\ref{fig:direct_power} (g)(h) demonstrate that although circuit optimization improves all the circuit quality, the intrinsic advantage of the direct methods in this scenario still persists. In Fig.~\ref{fig:direct_power} (e)(f)(g)(h), PSR, DHT and DRHT achieve the best quality while RHT and RDHT suffer from additional SWAP gates for interacting with the ancillary qubit. The key observation is that non-direct methods rely on the ancillary qubit and controlled multi-qubit gates, for generators that span over many qubits, all those qubits need to interact with the ancillary qubit. Thus the SWAP (CNOT) overhead of those algorithms severely undermines the circuit fidelity. Besides the quality analysis, we also perform end-to-end training with different gradient methods on the 6-qubit scale to verify that good circuit quality helps with training convergence. One can see from Fig.~\ref{fig:direct_power} (i) that even though direct methods have $2\times$ cost per iteration, the quality gap is significant enough to overcome that cost, and direct methods converge substantially faster compared to their counterparts.

\remark{We look forward to future work that integrates other circuit optimization techniques with our gradient framework. We anticipate that as more powerful optimizers emerge, the overall circuit quality will improve, thereby narrowing the quality gap between different gradient methods. Consequently, the convergence rate is expected to rely more heavily on the quantity of circuits, quantified by the distinct circuit execution count. Our reversed methods are designed specifically to address this aspect. In QAD, the balance between quality and quantity can be easily adjusted based on user preferences, as illustrated in Table.~\ref{tab:models}.}